\documentclass[notitlepage,nofootinbib,twocolumn,amsmath]{revtex4-1}

\usepackage{amsfonts}
\usepackage{amsmath}
\usepackage{amssymb}
\usepackage{aas_macros}
\usepackage{graphicx}
\usepackage{color}
\usepackage{float}
\allowdisplaybreaks[1]

\def\lsim{~\rlap{$<$}{\lower 1.0ex\hbox{$\sim$}}}
\def\bsim{~\rlap{$>$}{\lower 1.0ex\hbox{$\sim$}}}

\def\hkpc{\ {\rm {\it h}^{-1}Kpc}}
\def\hmpc{\ {\rm {\it h}^{-1}Mpc}}

\def\hmsun{\ {\rm {\it h}}^{-1}M_\odot}

\def\cmmm{\ {\rm cm^{-3}}}
\def\cmm{\ {\rm cm^{-2}}}
\def\masslength{\ {M_\odot \,{\rm Mpc}^{-1}}}
\def\surfdens{\ {h\, M_\odot \,{\rm Mpc}^{-2}}}

\def\hmmpc{\ {\rm {\it h}Mpc^{-1}}}

\def\eV{\ {\rm eV}}
\def\GeV{\ {\rm GeV}}
\def\Kel{\ {\rm K}}

\def\apjl{Astrophys.\ J.\ Lett.}

\def\aap{Astron.\ Astrophys.}

\def\apjs{Astrophys.\ J. Supp.}

\def\ln{{\rm ln}}

\def\mathbi#1{\textbf{\em #1}}

\def\rvh{\mathrm{\hat{\bf{r}}}}

\def\vk{\mathbi{k}}

\def\vr{\mathbi{r}}

\def\vu{\mathbi{u}}

\def\vx{\mathbi{x}}

\def\grad{\boldsymbol{\nabla}}

\definecolor{RedWine}{rgb}{0.743,0,0}
\definecolor{RoyalBlue}{rgb}{0.25,.41,.88}
\definecolor{ForestGreen}{rgb}{.13,.54,.13}
\definecolor{DeepPurple}{rgb}{.72,.18,1}

\begin{document}

\title[The impact of ultra-light axion self-interactions on the large scale structure of the Universe]
      {The impact of ultra-light axion self-interactions \\ on the large scale structure of the Universe}

\author{Vincent Desjacques$^1$}
\email{dvince@physics.technion.ac.il}
\affiliation{$^1$Physics department and Asher Space Science Institute, Technion, Haifa 3200003, Israel}

\author{Alex Kehagias$^2$}
\email{kehagias@central.ntua.gr}
\affiliation{Physics Division, National Technical University of Athens, 15780 Zografou Campus, Athens, Greece}

\author{Antonio Riotto$^3$}
\email{Antonio.Riotto@unige.ch}
\affiliation{Department of Theoretical Physics and Center for Astroparticle Physics (CAP) 24 quai E. Ansermet, CH-1211 Geneva 4, Switzerland}


\date{\today}
\label{firstpage}

\begin{abstract}
\noindent
Ultra-light axions have sparked attention because their tiny mass $m\sim 10^{-22}$ eV, which leads to a Kiloparsec-scale de Broglie wavelength
comparable to the size of dwarf galaxy, could alleviate the so-called small-scale crisis of massive cold dark matter (CDM) candidates.
However, recent analyses of the Lyman-$\alpha$ forest power spectrum set a tight lower bound on their mass of $m\gtrsim 10^{-21}$ eV which
makes them much less relevant from an astrophysical point of view. An important caveat to these numerical studies is that they do not take
into account self-interactions among ultra-light axions. Furthermore, for axions which acquired a mass through non-perturbative effects, this
self-interaction is attractive and, therefore, could counteract the quantum ``pressure'' induced by the strong delocalization of the particles.
In this work, we show that even a tiny attractive interaction among ultra-light axions can have a significant impact on the stability of cosmic
structures at low redshift. After a brief review of known results about solitons in the absence of gravity, we discuss the stability of
filamentary and pancake-like solutions when quantum pressure, attractive interactions and gravity are present. 
The analysis based on one degree of freedom, namely the breathing mode, reveals that pancakes are stable, while filaments are unstable if the
mass per unit length  is larger than a critical value. However, we show that  pancakes are unstable against transverse perturbations. We expect
this to be true for halos and filaments as well.
Instabilities driven by the breathing mode will not be seen in the low column density Lyman-$\alpha$ forest unless the axion decay constant
is extremely small, $f\lesssim 10^{13}\GeV$. Notwithstanding, axion solitonic cores could leave a detectable signature in the Lyman-$\alpha$
forest if the normalization of the unknown axion core - filament mass relation is $\sim 100$ larger than it is for spherical halos.
We hope our work motivates future numerical studies of the impact of axion self-interactions on cosmic structure formation.
\end{abstract}

\maketitle

\section{Introduction}
\label{sec:intro}

The idea that the dark matter in our universe could be formed of ultra-light bosons can be traced back to the work of \cite{Baldeschi:1983mq,Sin:1992bg}.
Lately, it has attracted a lot of attention owing to the fact that, for a particle of mass around $10^{-22}$ eV, the corresponding de Broglie wavelength,
which defines the scale at which ``quantum pressure'' sets in,  is about a Kpc. Therefore, this would alleviate the small-scale problems of the cold dark
matter candidates \cite{viel1,viel2,viel4,viel7} (for a review, see Ref. \cite{viel14}).
The cosmological properties of such ultra-light bosons have been scrutinized in details, from the characterization of the linear power spectrum
\cite{viel1,viel12}  to the numerical analysis of the non-linearities at small scales through in N-body simulations \cite{viel10}, the study of the
innermost structure of halos \cite{viel8},  the dynamical properties of the smallest objects \cite{viel9} and the impact on galaxy formation \cite{Mocz:2017wlg}.  

On the other hand, the hypothesis of ultra-light bosons as dark matter has been recently challenged in  a series of papers based on measurements of the
Lyman-$\alpha$ forest power spectrum extracted from high-redshift quasars \cite{realviel,marshalpha}.
The Lyman-$\alpha$ forest arises from the filamentary and sheet-like nature of the highly ionized, high-redshift intergalactic medium (IGM).
It probes fluctuations in the matter distribution on scales $k\gtrsim 1\hmmpc$ \cite{Croft:1997jf,Croft:1998pe,McDonald:1999dt,viel19} and, therefore, is
very sensitive to properties of the dark matter such as its free-streaming scale \cite{Viel:2005qj,Seljak:2006qw}.
Fluctuations in the Lyman-$\alpha$ forest set stringent lower bounds on the mass of a ultra-light bosons, $m > 2\times 10^{-21}\eV$ (95\% C.L.)
\cite{realviel,marshalpha}, which appear to significantly limit the role of ultra-light bosons in cosmology (see also \cite{visinelli:2017}).
By contrast, Ref. \cite{Zhang:2017chj} argues that $m=10^{-22}\eV$ is still consistent with the data owing to uncertainties in the thermal state of the
high-redshift IGM. 

The goal of this paper is to offer the first  study of the impact of self-interactions among the ultra-light bosons on the
mildly nonlinear large scale structure as traced by the Lyman-$\alpha$ forest.
In particular, we will investigate the impact of an attractive force induced by a quartic coupling on filamentary and sheet-like structures.
Such an attractive force arises when ultra-light bosons are identified with ultra-light CP-odd axions, which arise from a symmetry-breaking conjectured
  to solve the strong CP problem \cite{Weinberg:1977ma,Wilczek:1977pj,PQ1,PQ2}.
For a mass of the order $m\sim 10^{-22}$ eV and a decay constant (or symmetry-breaking scale) $f\sim 10^{17}$ GeV,
these ultra-light axions may provide a large fraction of the dark matter component as the energy of its oscillating condensate contributes a fraction
(see, e.g., \cite{viel14})
\begin{equation}
\Omega\sim 10^{-1}\left(\frac{f}{10^{17}\, {\rm GeV}}\right)^2\left(\frac{m}{10^{-22}\, {\rm eV}}\right)^{1/2}
\end{equation}
to the present-day critical density. 
For these fiducial values the corresponding  quartic coupling is extremely tiny, 
\begin{equation}
\lambda=\frac{m^2}{f^2}\sim 10^{-96}\;,
\end{equation}
and, at first sight, completely negligible. This is the reason why nearly all studies in the literature set this self-coupling constant to zero.
However, despite its tininess, the attractive self-interaction of the ultra-light axions plays a crucial role.

To convince oneself about the importance of the small attractive forces among the axions, consider the case of the spherical three-dimensional halos
made of axions. In the absence of gravity, such halos are always unstable. In the presence of gravity, spherical halos  are  stable only if their
masses are smaller than about 
\begin{equation}
M\sim 7\times 10^{9}\,\hmsun \quad \mbox{for} \quad
\lambda\sim 10^{-96}\;. 
\end{equation}
This has been known since the seminal work of Vakhitov and Kolokolov \cite{vv} and  stressed again more recently in
Refs.~\cite{Chavanis:2011zi,chavanis/delfini:2011,eby/suranyi/etal:2015,chavanis:2016,Levkov:2016rkk,eby/ma/etal:2017,helfer/marsh/etal:2017}.
This result can be easily understood if one realizes that the effective self-interaction coupling is not $\lambda$ itself, but
$\lambda$ multiplied by the phase-space density of axions in the environment. Since there are situations in which the latter is huge, the attractive
force may become important.  This phenomenon is well-known in non-linear physics as it is responsible for the self-focusing of laser beams for instance.
Consequently, the current small-scale results extracted from N-body simulations in which the quartic coupling has been dismissed, so that there is no
critical mass above which halos are unstable, should be reconsidered.

As we already mentioned however, the fundamental objects giving rise to the Lyman-$\alpha$ forest used to set stringent bounds on the mass of the
ultra-light axions are the IGM filaments and pancakes.
Therefore, the following question naturally arises: do these structures exist when the dark matter is composed by ultra-light axions ?
More generally, in light of the instability of massive halos, it is desirable to investigate how the cosmic web looks like when the Universe is dominated
by a sizeable fraction of self-attracting ultra-light bosons.

Our analytical findings indicate that the cosmic web is influenced by a small, non-vanishing self-coupling among ultra-light axions.
In particular, pancakes are unstable against transverse perturbations even in the presence of gravity; filaments are unstable if their mass per unit length
is larger than some critical value owing to the increase of the attractive axion self-interaction which causes the filaments to eventually collapse.
These results indicate that a more thorough investigation should be performed at the numerical level in order to properly assess whether ultra-light axions
are ruled out by Lyman-$\alpha$ forest data and, on a broader scope, to understand cosmic structure formation in this scenario.
Most cosmological simulations of ultra-light axions thus far have ignored axion self-interactions, focusing mainly on the impact of their large de Broglie
wavelength \cite{woo/chiueh:2008,viel10,veltmaat/niemeyer:2016,zhang/etal:2016}.

The paper is organized as follows. In \S\ref{sec:linear}, we estimate the impact of the attractive force among axions in the linear regime.
In \S\ref{sec:nograv_stability}, we study the stability of the cosmic web beyond the linear order in the case in which gravity is switched off.
This section contains results well-known in the non-linear physics community exploring the properties of Bose-Einstein condensates.
\S\ref{sec:grav_stability} is devoted to the stability analysis when gravity is turned on, whereas \S\ref{sec:lymanalpha} investigate the possibility of
detecting unstable axion filaments in the Lyman-$\alpha$ forest. Finally, \S\ref{sec:conclusions} summarizes our findings and conclusions.
We use natural units $\hbar=c=k_B=1$ throughout, and adopt a cosmological model consistent with CMB data \cite{Hinshaw:2012aka,Ade:2015xua}.

\section{Axion perturbations in the linear regime}
\label{sec:linear}

Our starting point is the action for the ultra-light axion field $\phi$,
\begin{equation}
S[\phi]=\int\! d^4x\,\sqrt{-g}\left[\frac{1}{2}(\partial\phi)^2-\Lambda^4\left(1-\cos\frac{\phi}{f}\right)\right]\;,
\end{equation}
where $\Lambda$ is a sort of condensation scale, $f$ is the decay constant and $g$ is the metric determinant.
Expanding for $\phi\ll f$ and including the quartic coupling, one obtains a potential of the form
\begin{gather}
  \label{eq:Vaxion}
  V(\phi) = \frac{1}{2} m^2 \phi^2 - \frac{1}{4!} \lambda \phi^4 \;, \\
  \mbox{where} \quad  m^2=\frac{\Lambda^4}{f^2} \quad \mbox{and} \quad\lambda = \frac{m^2}{f^2} \;. \nonumber
\end{gather}
For our fiducial choices of axion mass $m=10^{-22}\eV$ and decay constant $f=10^{17}\GeV$, we find 
$\lambda=+10^{-96}$. Higher order terms ($\phi^6$ and higher) are negligible as long as $\phi/f\ll 1$ and this remains true as well when 
taking into account the high phase-space density.
Notice that $\phi$ has dimension of energy, and that the sign of the self-interaction coupling leads to an 
attractive force. This will be relevant for all our considerations.

The cross-section per unit mass is
\begin{equation}
\frac{\sigma}{m} = \frac{\lambda^2}{32\pi m^3} \sim 10^{-97} \mbox{cm$^2$/g} \;.
\end{equation}
For comparison, constraints on self-interacting dark matter from merging of galaxy clusters impose the upper bound $\sigma/m\lesssim 1$ cm$^2$/g
\cite{Randall:2007ph}. 
Therefore, one would naively expect that axion self-interaction is completely negligible as far as astrophysical scales are concerned. 
As we shall see later however, because the axion phase space density is enormous, self-interaction can play a role at sufficient large number
densities.

We will now perform a stability study and assess the relevance of the axion self-interaction at the linear level.
The linear analysis is discussed in Ref. \cite{Harko:2011zt,Chavanis:2011uv}.
To derive the non-relativistic limit of the Klein-Gordon equation 
\begin{equation}
\label{eq:KG}
\Box\phi -m^2 \phi =- \frac{\lambda}{3!} \phi^3\;, 
\end{equation}
we set
\begin{equation}
\phi(\vx,\eta) = \sqrt{2}{\rm Re}\left(\psi(\vx,\eta)\, e^{-im\int d\eta' a'}\right) \;,
\end{equation}
where we have adopted the conformal time $\eta$ and the complex phase of $\psi(\vx,\eta)$ is such that
\begin{equation}
-i(E-m)\int\! d\eta' a' \ll -im\int\! d\eta' a'
\end{equation}
in the 
non-relativistic limit.
Note that $\psi$ has the same units as $\phi$, i.e. units of energy.
We thus obtain
\begin{align}
\dot{\phi}(\vx,\eta) &= \sqrt{2}\,{\rm Re}\Big[\left(\dot{\psi}-ima \psi\right) e^{-im\int\! d\eta' a'}\Big] \\
&\approx -\sqrt{2}\,{\rm Re}\left(ima\psi\, e^{-im\int\! d\eta' a'}\right) \nonumber \;,
\end{align}
and
\begin{align}
\ddot{\phi}(\vx,\eta) &= 
\sqrt{2}\,{\rm Re}\bigg\{\Big[\ddot{\psi}-im\left(2a\dot{\psi}+\dot{a}\psi\right)- m^2 a^2\psi\Big] \\
& \qquad \times e^{-im\int\! d\eta' a'}\bigg\} 
\nonumber \\
&\approx -\sqrt{2}\,{\rm Re}\bigg\{\Big[im\left(2a\dot{\psi}+\dot{a}\psi\right)+m^2 a^2 \psi\Big] \nonumber \\
& \qquad \times e^{-im\int\! d\eta' a'}\bigg\} \nonumber \;.
\end{align}
In each expression, we have neglected the term with highest time-derivative, as it is strongly suppressed relative to the others.
Substituting these relations into the Klein-Gordon equation, we arrive at the Gross-Pitaevskii-Poisson (GPP) equations in the
non-relativistic Newtonian gauge (upon averaging over the fast period set by the axion mass)
\begin{align}
i a\bigg(\partial_\eta\psi+\frac{3}{2}\mathcal{H}\psi\bigg) &= 
-\frac{1}{2m} \Delta_\vx\psi + m a^2\left(\Phi - \frac{1}{8f^2}|\psi|^2\right)\psi \nonumber \\
\Delta_\vx\Phi &= \frac{4\pi}{m_P^2}a^2\rho \;,
\end{align}
where $\Phi$ is the Newtonian gravitational potential. We have included in the axion energy density the dominant piece solely, that 
is, $\rho \approx  m^2 |\psi|^2$.
Ref. \cite{Levkov:2016rkk} proposes a coordinate and field rescaling that absorbs all physical constants. 
However, this rescaling involves explicitly $m_P$ and, thus, is not suited to study the limit in which the gravitational interaction
becomes negligible. Therefore, we decided to rescale the coordinates and the fields according to
\begin{gather}
\eta \to \frac{1}{m}\eta=\tilde \eta\;,\quad \vx=\tilde \vx\;, \quad 
\psi\to \frac{m}{f}\psi=\tilde \psi\;, \\  \rho\to \frac{1}{f^2}\rho = \tilde \rho\;, \quad 
\Phi\to m^2 \Phi=\tilde \Phi \nonumber \;.
\end{gather}
The system of equations can be recast into
\begin{align}
ia\bigg(\partial_\eta\psi+\frac{3}{2}\mathcal{H}\psi\bigg) &= 
-\frac{1}{2}\Delta_\vx\psi + a^2 \left(\Phi - \frac{1}{8}|\psi|^2\right) \psi 
\label{eq:GPPnodim} \\
\Delta_\vx\Phi & = 4\pi \tilde G a^2 |\psi|^2 \nonumber \;,
\end{align}
where the gravitational constant $\tilde G=(m f)^2/m_P^2$ has dimensions of energy square, and we have dropped the 
tildes from the coordinates and the fields to avoid clutter.
We perform the standard  Madelung transformation and write the wave function $\psi(\vx,\eta)$ as
\begin{equation}
\psi(\vx,\eta) = A(\vx,\eta) e^{i\theta(\vx,\eta)} \;.
\end{equation}
Since the density now reads $\rho = |\psi|^2 \equiv A^2$,
the normalized GGP system Eq.~(\ref{eq:GPPnodim}) can be recast into the form
\begin{align}
\dot{\rho}+3\mathcal{H}\rho + \grad_\vx\left(\rho\vu\right) &= 0\; , \\
\dot{\vu}+\mathcal{H}\vu+\left(\vu\cdot\grad_\vx\right)\vu &= - \grad_\vx Q - \grad_\vx\Phi-\grad_\vx h\; , 
\nonumber \\
\Delta_\vx\Phi &= 4\pi\tilde G a^2\rho \nonumber \;,
\end{align}
upon defining the axion phase velocity $\vu\equiv a^{-1}\grad_\vx\theta$. This shows that $\vu(\vx,\eta)$ is the physical,
peculiar (bulk) velocity of the axion condensate. Furthermore,
\begin{align}
Q=-\frac{1}{2a^2}\frac{\Delta_\vx\sqrt{\rho}}{\sqrt{\rho}}
\end{align}
is the quantum potential and $h(\rho) = -\rho/8$ is the enthalpy per unit mass.

\begin{figure*}
\includegraphics[width=0.49\textwidth]{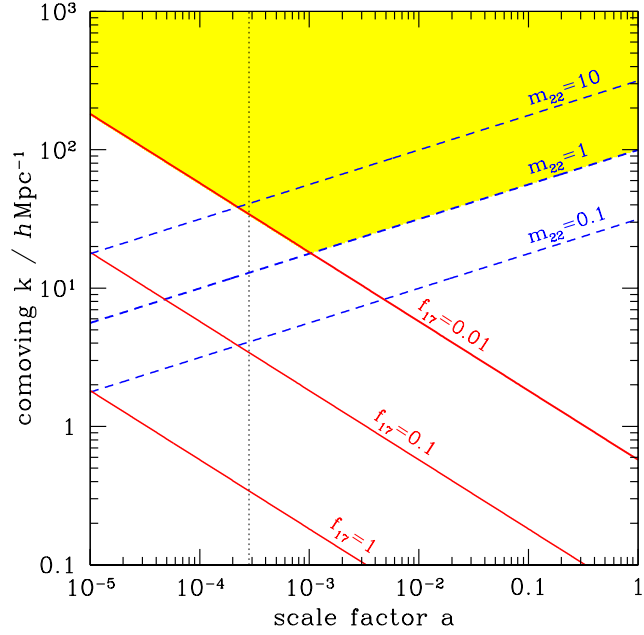}
\caption{The comoving wavenumber $k_J(a)$ (dashed green) and $k_I(a)$ (solid red) as a function of scale factor $a$ for 
different axion masses $m_{22}$ and decay constant $f_{17}$ (see text).
The shaded (yellow) area indicates the region that is linearly stable to perturbations, i.e. $k>$min$(k_J,k_I)$. 
The vertical (dotted) line indicate the scale factor $a_\text{eq}$ of equivalence ($z_\text{eq}=3515$ in our cosmology with
$\Omega_mh^2=0.147$). When the axion self-interaction is neglected, stability occurs above the dashed (blue) curve. }
\label{fig:scales}
\end{figure*}

We now  linearize the GPP system. The contribution from the so-called quantum ``pressure'' $Q$ (which purely arises from
the uncertainty principle: the delocalization of the particles increases with their momenta) is given by
\begin{align}
\grad_\vx Q &= -\frac{1}{2a^2}\grad_\vx\left(\frac{\Delta_\vx \sqrt{\rho}}{\sqrt{\rho}}\right) \\
&= -\frac{1}{2a^2}\grad_\vx\left(\frac{\Delta_\vx\sqrt{1+\delta}}{\sqrt{1+\delta}}\right)\nonumber \\
&\approx -\frac{1}{4a^2} \grad_\vx\left(\Delta_\vx\delta\right) \nonumber \;.
\end{align}
Similarly, the interaction term becomes
\begin{equation}
\grad_\vx h(\rho)=- \frac{1}{8} \grad_\vx\rho=-\frac{\bar\rho}{8}\grad_\vx\delta \;,
\end{equation}
where $\bar\rho$ is the physical, average density of axions.
Substituting these expressions into the Euler equation,  we obtain the linear growth equation
\begin{equation}
\ddot{\delta}+ \mathcal{H}\dot{\delta} + \frac{1}{4a^2}\Delta_\vx^2\delta + \frac{\bar\rho}{8}\Delta_\vx\delta
-4\pi\tilde G a^2\bar\rho\delta = 0 
\end{equation}
or, in Fourier space,
\begin{equation}
\ddot{\delta_\vk}+ \mathcal{H}\dot{\delta_\vk}+\left(\frac{k^4}{4 a^2}- \frac{\bar\rho k^2}{8}-4 \pi \tilde G a^2\bar\rho\right)
\delta_\vk = 0 \;,
\end{equation}
where $\delta_\vk$ is the amplitude of the Fourier modes. Going back to the dimensionful variables (physical units), this reads
\begin{equation}
\ddot{\delta_\vk}+\mathcal{H}\dot{\delta_\vk}+\left(\frac{k^4}{4 a^2 m^2}- \frac{\bar\rho k^2}{8m^2f^2}-4 \pi G a^2\bar\rho\right)
\delta_\vk = 0 \;.
\end{equation}
Like gravity, the self-interaction also induces a contribution proportional to the mean density $\bar\rho$.
Ignoring the self-interaction, the quantum pressure and gravitational pull define a characteristic (comoving) ``Jeans scale'' 
\begin{equation}
\label{eq:kJ}
k_J = (16\pi G)^{1/4} m^{1/2} a \bar\rho^{1/4} \;.
\end{equation}
For non-relativistic dust with $\bar\rho \propto a^{-3}$ and a Hubble parameter $h=0.7$, the comoving Jeans scale is given by
\begin{equation}
k_J(a) = 161\, a^{1/4} m_{22}^{1/2}\left(\Omega_m h^2\right)^{1/4} \hmmpc
\end{equation}
where, for convenience, we shall work with 
\begin{align}
m_{22} &\equiv \frac{m}{10^{-22}\eV} \;, \\
f_{17} &\equiv \frac{f}{10^{17}\GeV} \;, \nonumber \\
\lambda_{96} &\equiv \frac{\lambda}{10^{-96}} \nonumber \;.
\end{align}
In configuration space, the corresponding Jeans length is
\begin{equation}
r_J(a) = 2\pi/k_J = 39\, a^{-1/4} m_{22}^{-1/2}\left(\Omega_m h^2\right)^{-1/4} \hkpc \;.
\end{equation}
in agreement with \cite{Hu:2000ke}.

Similarly, the self-interaction becomes larger (in magnitude) than the quantum pressure at wavenumbers $k<k_I$, where
\begin{equation}
k_I =  2^{-1/2} a f^{-1} \bar\rho^{1/2} \;.
\end{equation}
This corresponds to the (comoving) characteristic wavenumber
\begin{equation}
  \label{eq:kI}
k_I(a) = 1.5\times 10^{-2} a^{-1/2} f_{17}^{-1} \left(\Omega_m h^2\right)^{1/2} \hmmpc
\end{equation}
which, in configuration space, translates into the 
\begin{equation}
r_I(a) = 68\, a^{1/2} f_{17} \left(\Omega_m h^2\right)^{-1/2} \hmpc \;.
\end{equation}
In other words, the self-interaction always dominates the quantum pressure on scales $k\lesssim 1\hmmpc$ for our fiducial 
parameter values. 

The characteristic wavenumbers $k_J$ and $k_I$ are shown in Fig.\ref{fig:scales} as a function of scale factor for various
choices of $m_{22}$ and $f_{17}$. Perturbations with a comoving wavenumber $k$ are linearly stable when they lie within the
shaded area. In linear theory, the axion self-interaction is relevant at high redshift, but completely negligible at
low redshift. However, it affects the growth of perturbation only in radiation domination so long as the decay constant is
$f_{17}\gtrsim 0.03$.
Therefore, this suggests that axion self-interaction can be safely neglected in the linear regime if all the dark matter is
in the form of axions.

Let us conclude this Section with a brief discussion of \cite{Cedeno:2017sou,bump1,bump2} (and the numerical study of
\cite{schive/chiueh:2018}). These authors implemented the
full axion potential $\Lambda^4\big(1-\cos(\phi/f)\big)$ into a {\small CMB} Boltzmann code and found that,
for $f_{17}\lesssim 0.1$ (which corresponds to their coupling being $\gtrsim 10^5$),
the $z=0$ linear power spectrum exhibits a bump around $k\gtrsim 1\hmmpc$.
They attribute this feature to a tachyonic instability
~\footnote{The potential Eq.(\ref{eq:Vaxion}) is tachyonic ($V''<0$) for $\phi/f\geq \sqrt{2}$, i.e. outside the range of
validity of our approximation (which assumes $\phi\ll f$).}
of the linear mode.
This effect arises only when the initial misalignment angle $\theta_i\equiv\phi_i/f$ (where $\phi_i$ is the initial field
value) is very close to $\pi$, that is, $\phi$ starts at the top of its potential. Such highly fine-tuned initial conditions,
which imply $\phi_i\sim f$, will not be considered here.



\section{Stability analysis beyond the linear regime: excluding gravity}
\label{sec:nograv_stability}

In this section we start our considerations about the impact of the self-interactions of the ultra-light axions on the
nonlinear large scale structure. Albeit tiny, they can have a huge influence on the cosmic web (namely, halos, pancakes and
filaments).

The fundamental implications of an attractive force have been widely studied in non-linear physics.
They lead, for instance, to the phenomena of modulation instability and catastrophic self-focusing of laser beams or collapse in
Bose-Einstein condensates, with the  collapse  being sometimes self-similar,
i.e. described by mathematical solutions whose forms  are  rescaled ground-state solitary waves or solitons. 
We consider first the Gross-Pitaevskii (GP) equation without gravity  (and, therefore, ignore the expansion of the Universe).
The following results, which are extensively discussed in \cite{sulem1999nonlinear}, are standard in the condensed matter community.
However, we are of the idea that summarizing here the salient features is useful for the reader who might not be familiar with
these arguments\footnote{In all these considerations we assume that the number of particles is conserved.
  This of course fails to be true during a collapse phase when the self-annihilation of the axion field  into its own relativistic
  quanta and/or photons become relevant due to resonant effects, see for instance Ref.  \cite{Riotto:2000kh}.}.
  More details can be found in Ref. \cite{sulem1999nonlinear}.

\subsection{Solitons in $D$-dimensions}
\noindent
We begin with the elliptic, nonlinear GP equation in $D$-dimensions written in (cosmic) time $t$ and rescaled coordinates (notice that
we have appropriately rescaled  the spatial coordinates and the wave function in order to eliminate all the irrelevant coefficients)

\begin{equation} 
\label{el}
i \partial_t\psi+ \Delta_\vx\psi +  |\psi|^{2}\psi=0 \;.
\end{equation}
One can look for standing wave solutions of the form

\begin{equation}
\psi(\vx,t)=e^{i\omega t}\phi(\vx)\;,
\end{equation}
where the function $\phi(\vx)$ satisfies the equation

\begin{equation} 
 \Delta_\vx\phi -\omega\phi+  |\phi|^{2}\phi=0 \;,
\end{equation}
and $\omega$ has to be positive to ensure that the solution (and its derivatives) vanishes at spatial infinity.
One can easily prove that the solution for the field $\phi$ arises from the variational problem

\begin{equation}
\label{a}
\delta\left(H+\omega N\right)=0\;,
\end{equation}
where 

\begin{align}
  N &= \int\! d^D\vx\, |\phi|^2 \\
  H &= \int\! d^D\vx\,\left(|\nabla_\vx\phi|^2-\frac{1}{2} |\phi|^4\right) \nonumber \;.
\end{align}
Here, $\Omega=H+\omega N$ and $-\omega\leq 0$ formally are the grand-canonical and chemical potential of the system.
For simplicity however, we shall refer to $\Omega$ as an effective energy rather than a grand potential. Note that
the chemical potential is negative to allow for the number $N$ of particles to be arbitrary large.

Solitons form owing to the balance between non-linear interaction and dispersion due to the quantum pressure.
They correspond to stationary points of their Hamiltonians under the condition that the  integral of motion,
i.e. the particle number, is  kept fixed.
Solitons are therefore said to be associated to conditional extrema. This is an important point to stress.
Were the GP solitons only stationary points of the Hamiltonian, the latter  would be unstable for those cases
in which the Hamiltonian is unbounded from below\footnote{This is an essential argument in Derrick's theorem \cite{Derrick}
about the stability of solitonic solutions
in dimension $D\geq 3$.}.
For solitons which are {\it stationary  and conditional} points of the Hamiltonian (i.e. they minimize the effective energy $\Omega$),
one can invoke the Lyapunov theorem\footnote{According to this theorem,  at least one stable solution is present  when some integral,
  e.g. the Hamiltonian,
  is  bounded.  Consider for instance  a system in a  state corresponding to the absolute minimum of such
  integral. Each variation of the solution must increase its value in contradiction to the integral conservation.
  Hence, the solution must be stable.}
so that, in order to prove the soliton stability, it is sufficient to prove the boundness of the Hamiltonian while the total
number of particles is held fixed.

The following transformation (a uniform stretching of the coordinates)

\begin{equation}
\widetilde\phi(\vx)=\frac{1}{a^{D/2}}\phi\left(\frac{\vx}{a}\right)
\end{equation}
preserves $N$ and transforms the Hamiltonian $H$ into

\begin{equation}
H(a)=\frac{1}{a^2}\int\! d^D\vx\,|\nabla_\vx\phi|^2-\frac{1}{2a^D}\int\! d^D\vx\,
|\phi|^4\;.
\end{equation}
Eq. (\ref{a}) then implies that 

\begin{equation}
\left.\frac{\partial H}{\partial a}\right|_{a=1}=0
\end{equation}
or

\begin{equation}
  H_1=\frac{D}{4}H_2 \;,
\end{equation}
where

\begin{equation}
H_1=\int\! d^D\vx\, |\nabla_\vx \phi|^2 \quad \mbox{and}\quad H_2=\int\! d^D\vx\, |\phi|^4
\end{equation}
are the kinetic and (minus) the potential energy, respectively.
On the other hand, since $H_1=-\omega N+H_2$ (the equation of motion implies that the energy $\Omega$ is zero for $a=1$), one obtains

\begin{eqnarray}
H_1&=&\frac{D\omega^{(4-D)/2}}{4-D}N_0\;,\nonumber\\
H_2&=&\frac{4\omega^{(4-D)/2}}{4-D}N_0\;,\nonumber\\
H&=&\frac{(D-2)\omega^{(4-D)/2}}{4-D}N_0\;,
\end{eqnarray}
where $N_0=\omega^{D-1}N$ parameterizes the total number of particles.  
From this variational principle argument, one sees that the Hamiltonian of the system evaluated at its ground state is positive for $D<2$
and vanishes at the critical dimension $D=2$. Furthermore, since 

\begin{equation}
\left.\frac{\partial^2 H}{\partial a^2}\right|_{a=1}=2(2-D)H_1\;,
\end{equation}
the ground state realizes a minimum of $H(a)$ when $D<2$ and a maximum when $D>2$.
One concludes that the standing wave which provides the ground state is stable for $D<2$ and unstable for $D>2$. 

The case $D=2$ is called critical: both the kinetic and the potential energy terms  in the Hamiltonian have a similar scaling, and the  above analysis
cannot furnish a conclusive answer about the stability of the ground state solitons.
If the kinetic energy is larger than the (minus) potential energy, the Hamiltonian  is positive, increasing  as $a$ tends to zero;
in the opposite case, the Hamiltonian  is unbounded from below; 
all solitonic solutions on the $(H,a)$-plane are degenerated into the line $H=0$, whose soliton represents a sort of separatrix between the manifolds
of collapsing and non-collapsing distributions.
Let us elaborate further on this. 
Consider the variance 

\begin{equation}
{\cal V}=\int\! d^2\vx \,r^2\,|\psi|^2\;,
\end{equation}
with $r=|\vx|$. ${\cal V}$ measures the width of the soliton if the latter is centered at the origin. 
Using the fact that

\begin{equation}
\frac{d^2{\cal V}}{dt^2} = \int\!d^2\vx\,r^2\, \partial_i\partial_jT_{ij} \;,
\end{equation}
where

\begin{align}
  T_{ij} &= 2\Big(\partial_i\psi\partial_j\psi^*+\partial_i\psi^*\partial_j\psi+{\cal L}\delta_{ij}\Big)\; , \\
  {\cal L} &= -\frac{1}{2}\Big\{\nabla_\vx\cdot\big(\psi^*\nabla_\vx\psi+\psi\nabla_\vx\psi^*\big)+|\psi|^4\Big\}
  \nonumber \;,
\end{align}
are the momentum stress tensor and the Lagrangian density.
Repeated integration by parts leads to the identity \cite{sulem1999nonlinear}

\begin{align}
  \frac{d^2{\cal V}}{dt^2} &= 2\int\!d^2\vx\,\sum_j T_{jj}  \\
  &= 8\left(\int\! d^2\vx\,|\nabla_\vx \psi|^2-\frac{1}{2}\int\! d^2\vx \,|\psi|^4\right) \nonumber \\
  &\equiv 8H \nonumber\;.
\end{align}
Since the Hamiltonian is conserved, this implies that

\begin{equation}
\label{aa}
{\cal V}=4 H t^2+ C_1 t+C_0\;,
\end{equation}
where the integrations constants $C_0$ and $C_1$ are related to additional conserved quantities.
Because ${\cal V}$ is positive by construction, if $H<0$ (called the Vlasov-Petrishchev-Talanov criterion \cite{vlasov}) then ${\cal V}$
reaches zero at some finite time; if $H>0$ then ${\cal V}$ grows indefinitely.
In other words, any perturbations of solitons that move the Hamiltonian away from zero will  lead  either to  collapse for $H<0$  or
to a dynamics in which collapse is forbidden \cite{weinstein}.
In this latter case, the system  completely spreads out because of dispersion, the non-linear interactions become insignificant and waves
become linear at large times with their amplitude  tending to zero.
The ground state soliton of the critical value $D=2$ is the Townes ground soliton \cite{Townes}, and has the property of having Hamiltonian
$H = 0$ associated to some critical number of axions $N_c$ (see below).
Solitons with a number of particles smaller than $N_c$ will then be stable (because the Hamiltonian scales like $H\propto (N_c-N)$)
thus confirming our findings in the previous section.
Let us elaborate further about the implications of this generic analysis for the cosmic web made out of ultra-light axions.

\begin{flushleft}
{\bf Halos} 
\end{flushleft}
\noindent
According to the discussion above,   halos, which correspond to $D=3$ standing waves,  must be  unstable (in the absence of gravity).
This matches of course the findings of Ref. \cite{Chavanis:2011zi}, where it was found that, in the limit of vanishing gravity, there are no
stable three-dimensional halos for ultra-light bosons. 
The wave-function becomes singular in a finite amount of time.
This phenomenon is generically called supercritical wave collapse and it causes a fast transfer of energy from the large to the small scales
with the wave function scaling around the singularity time $t_0$ as \cite{Kuznetsov:1986vj,Levkov:2016rkk}
 
\begin{align}
  \psi(r,t) &= \frac{1}{(t_0-t)^{1/2+i\alpha}}\chi\left(\frac{r}{(t_0-t)^{1/2}}\right) \\
  \chi(\xi) &= C/\xi^{1+2i\alpha}\,\,\,{\rm for}\,\,\, \xi\gg 1 \nonumber \;,
\end{align}
where $\alpha\simeq 0.545$ and $C\simeq 1.01$.
When gravity is turned on again, it can help stabilizing the halos so long as they are smaller than a maximal mass, generating thereby what is
called a self-bound condensate.

\begin{flushleft}
{\bf Pancakes} 
\end{flushleft}
\noindent
Pancakes correspond to $D=1$ solitons and, therefore, are  stable according to the variational principle arguments. 

\begin{flushleft}
{\bf Filaments} 
\end{flushleft}
\noindent
As we have seen previously, for filaments, i.e. $D=2$ standing waves, the variational principle is inconclusive. 
In fact, one can show again that, at the critical dimension $D=2$, the wave-function may become infinite in a finite amount of time if filaments
are associated to a number of particles larger than a critical value $N_c$.
This phenomenon is generically called critical wave collapse. To find $N_c$ one may consider the $D=2$ GP equation 
 
\begin{equation} 
\label{eq:D2GP}
  i \partial_t\psi+ \left(\frac{\partial^2}{\partial x^2}+\frac{\partial^2}{\partial y^2}\right)\psi +  |\psi|^{2}\psi=0 \;.
\end{equation}
It gives rise to the so-called Townes solitons \cite{Townes}
 
\begin{equation}
\psi(x,y,t)=e^{i\omega t}\phi_{\rm T}(r)\;, \qquad r^2=x^2+y^2 \;,
\end{equation}
where $-\omega$ is an arbitrary chemical potential and $\psi$ satisfies the equation

\begin{equation}
\left(\frac{d^2}{d r^2}+\frac{1}{r}\frac{d}{d r}-\omega\right)\phi_{\rm T}+\phi_{\rm T}^3=0\;.
\end{equation}
The well-known  Vakhitov-Kolokolov stability criterion \cite{vv} $dN/d(-\omega)<0$, which reflects the fact that the number of particles should increase
when the chemical potential is lowered, tells us that instability is reached for filaments with particle number larger  than  

\begin{equation}
N_c=\pi\int_0^\infty\! dr\,r\, \phi_{\rm T}^2(r)\simeq 5.85\;.
\end{equation}
The Vakhitov-Kolokolov stability criterion does not hold for the Townes solitons themselves, which indeed are  degenerate as they all satisfy $N=N_c$.
As we already mentioned, Townes solitons separate $D=2$ solitons which are doomed to collapse if the associated number of particles
is too large ($N>N_c$), from those who can survive if their associated number of particles is small enough ($N<N_c$).
From Eq. (\ref{aa}) one sees that the  characteristic size  of the collapsing filaments  scales roughly like $(t_0-t)^{1/2}$
(up to logarithmic corrections, see later). 
Townes solitons for which $N=N_c$ are unstable themselves.
The instability is fully non-linear, i.e. it cannot be  described  by any  eigenvalue in the spectrum evaluated around the solitary wave and leading to an
instability.
In fact, the two eigenvalues $\lambda_\pm$ of the $D=2$ GP equation linearized at a solitonic solution exactly vanish (while, for $N>N_c$, they emerge on the
real axis, which generates a linear instability).
This corresponds to the fact that the $D=2$ GP equation has conformal symmetry which allows to conclude that, if $\psi(r,t)$ is a solution,
so is $\ell^{-1}\psi(r/\ell,t/\ell^2)$ \cite{t}.
In such a situation, there exists a self-similar solution near the singularity of the form \cite{sulem}

\begin{equation}
  \psi(r,t) \simeq \frac{1}{\ell}\chi(\xi) e^{i\tau +i \ell\xi^2/4}\;,
\end{equation}
with
\begin{gather}
\xi = r/\ell\;,\quad \tau=\int_0^t\!\frac{dt'}{\ell^2(t')}\; \\
 \ell(t) = \left(2\pi\frac{t_0-t}{\ln\,\ln (t_0-t)^{-1}}\right)^{1/2} \nonumber \;.
\end{gather}

\subsection{Solitons and the role of transverse instabilities}

So far we have summarized the stability landscape for solitons living in $D=1,2$ and 3.
A natural question is the fate of such objects when they are embedded in three dimensions. Let us consider, for example, pancake-like objects.  
It is well-known that they are unstable when immersed in higher dimensions, for instance against long-wavelength transverse fluctuations \cite{zakharov}.
Therefore, let us consider the $D=2$ GP equation, Eq.~ (\ref{eq:D2GP}), with a perturbation along one extra transverse direction \cite{sulem1999nonlinear}.
Since  the unperturbed soliton solution  reads (with $\omega>0$)

\begin{equation}
\psi_0(x,t)=\Psi(x) e^{i\omega t}\;,\,\,\,\Psi(x)=\frac{\sqrt{2\omega}}{{\rm cosh}(\sqrt{\omega}x)}\;,
\end{equation}
one looks for solution of the perturbed equation where there are tiny disturbances of the amplitude and the phase

\begin{equation}
\psi=\Psi(1+\chi)e^{i(\omega t+\rho)}\;.
\end{equation}
Expanding for small $\chi$ and $\rho$ we get 
\begin{equation}
\psi=\psi_0 +(f+ig) e^{i\omega t}\;,
\end{equation}
with $f=\chi\Psi$ and $g=\rho\Psi$.
Linearizing the problem leads to the following system of equations

\begin{align}
\left(\frac{d^2}{d x^2}-\omega+3\Psi^2\right)f&=\partial_t g- \frac{\partial^2 f}{\partial y^2}\;, \\
\left(\frac{d^2}{d x^2}-\omega+\Psi^2\right)g&=-\partial_t f- \frac{\partial^2 g}{\partial y^2}\;. \nonumber
\end{align}
One introduces now the slow variables $Y=\epsilon y$ and $T=\epsilon t$ and expand the functions
$f$ and $g$ in powers of $\epsilon$, $f=f_0+\epsilon f_1+\epsilon^2 f_2+\cdots$ and similarly for $g$. At leading order
one finds

\begin{align}
\label{eq:D1GP_0}
\left(\frac{d^2}{d x^2}-\omega+3\Psi^2\right)f_0&=0\;, \\
\left(\frac{d^2}{d x^2}-\omega+\Psi^2\right)g_0&=0\;, \nonumber
\end{align}
so that

\begin{align}
f_0&= a(Y,T)\Psi'(x)\;, \\
g_0&= b(Y,T)\Psi(x)\;. \nonumber
\end{align}
The physical interpretation of the functions $a$ and $b$ is a slow, long-wavelength modulation of both the amplitude and phase of the solitonic
solution, $\psi\simeq \Psi(x+a(X,T))e^{i(\omega t+b(Y,T))}$. At linear order in $\epsilon$ one gets

\begin{align}
\left(\frac{d^2}{d x^2}-\omega+3\Psi^2\right)f_1&=\partial_T g_0\;, \\
\left(\frac{d^2}{d x^2}-\omega+\Psi^2\right)g_1&= -\partial_T f_0\;, \nonumber
\end{align}
which yields

\begin{align}
f_1&= \partial_T b\frac{d\Psi}{d\omega}\;, \\
g_1&= -\frac{1}{2}\partial_T a x\Psi\;. \nonumber
\end{align}
At second order in $\epsilon$, one finds

\begin{align}
  \label{eq:D1GP_2}
\left(\frac{d^2}{d x^2}-\omega+3\Psi^2\right)f_2&=\partial_T g_1-\partial_{YY}f_0\;, \\
\left(\frac{d^2}{d x^2}-\omega+\Psi^2\right)g_2&=-\partial_T f_1-\partial_{YY}g_0\;. \nonumber
\end{align}
For a non-trivial solution $f_2$ to exist, this system must satisfy the solvability conditions\footnote{To understand the origin of this conditions,
  let us re-express the first line of Eq. (\ref{eq:D1GP_0}) as $L f_0=0$, where the differential
  operator $L$ is self-adjoint. Similarly, the first line of Eq. (\ref{eq:D1GP_2}) can be recast into the system $L f_2 = s$, where $s$ is the inhomogeneous
  term. Therefore, interpreting the integral of $s\cdot f_0$ as the scalar product $(s,f_0)$, we obtain $(s,f_0)=(L f_2,f_0)=(f_2,L f_0)=(f_2,0)\equiv 0$,
which is precisely the first relation in Eq. (\ref{eq:solvable}).}

\begin{align}
  \label{eq:solvable}
\int_{-\infty}^\infty\! dx\, f_0\,\partial_T g_1&= \int_{-\infty}^\infty\! dx\, f_0\,\partial_{YY} f_0\;, \\
\int_{-\infty}^\infty\! dx\, g_0\,\partial_T f_1&= -\int_{-\infty}^\infty\! dx\, g_0\,\partial_{YY} g_0 \nonumber
\end{align}
or, equivalently, 

\begin{align}
\partial_{TT}a&= \frac{4}{3}\,\omega\,\partial_{YY} a\;, \\
\partial_{TT}b &= -4\,\omega\,\partial_{YY} b\;. \nonumber
\end{align}
Since $\omega$ is positive, these conditions imply that the pancake-like solution is always unstable against the transverse long-wavelength perturbations.
In fact, in Ref. \cite{janssen} it was subsequently shown that there is  a critical value of  transverse wavelength above   which the soliton is unstable
against fluctuations which are not limited to be long-wavelength. 
We do not expect gravity to change the situation, as we will see in the next section.
 
The experience with pancakes immersed in a higher-dimensional set-up teaches us that the stability of the various solitons is not at all guaranteed.
It is generally accepted that the focusing  GP equation (similar to the one we discussed here with attractive self-interaction) does not have stable (bright)
solitons in $D=3$.
The reason is that  the quantum pressure is not enough to counteract the internal energy of the soliton.
The self-interaction energy indeed scales like $1/R^{\overline D}$, where ${\overline D}$ is the co-dimension of the soliton and $R$ its typical size
(see the next section for more details), whereas the quantum pressure scales like $1/R^2$.
Therefore, only for pancakes ($\overline D=1$) the quantum pressure can compensate the attractive self-interaction, and hence only the pancake can be stable in the
absence of gravity.
However, this stability concerns only the breathing mode, i.e. the mode that shares the same symmetries as the soliton itself.
In other words, the pancake is stable for planar symmetric fluctuations.
We have seen that arbitrary transverse fluctuations make the soliton unstable.
Therefore, in the absence of gravity, all solitons  are expected to be unstable in $D=3$
\footnote{We thank D.J.  Frantzeskakis and T.P. Horikis for discussions about this point.}.

All these preliminary considerations show that the cosmic web of the ultra-light axions is expected to be quite different from the one in the standard cold
or warm dark matter (CDM and WDM) scenario because there is no attractive interaction in this case.
In the next section, we bring gravity back into the game.
This will necessarily limit our capability in investigating the stability of the cosmic wave and we will restrict ourselves to a one-degree of freedom analysis,
that is, to the breathing radius mode.

\section{Stability analysis beyond the linear regime: including gravity}
\label{sec:grav_stability}

The goal of this section is to investigate the stability of structures making up the cosmic web, such as halos, pancakes and filaments, formed by the
axion dark matter in the presence of an attractive force plus gravity. As we have seen already in the previous section, the impact of the self-interaction
among axion particles is not at all negligible.

To get insights about the stability issue, we simplify the problem by reducing it to a one-degrees of freedom by considering fully symmetric solitonic
objects with various co-dimensions $\overline D$.
Namely, halos have zero-dimension, and therefore $\overline D=3$; pancakes are two-dimensional objects and therefore $\overline D=1$;
filaments are one-dimensional strings and, correspondingly, $\overline D=2$. Note that the case $\overline D=3$ was studied in detail in \cite{Chavanis:2011zi}.
This simplification allows us to include gravity into the stability analysis.
The goal of this section is to understand if the generic results described in the previous section hold in the presence of gravity.

\subsection{Hamiltonian}

For simplicity, we set the chemical potential $\omega$ to zero, that is, we leave the number of particles $N$ unconstrained.
Therefore, the fundamental quantity we look at is the Hamiltonian $H$ associated with the GPP system 
\begin{equation}
H = E_K + E_Q + U + W \;,
\end{equation}
where $E_K$ is the ``classical'' kinetic energy, $E_Q$ is the quantum pressure, $U$ is the internal energy 
and $W$ is the gravitational energy. Namely,
\begin{gather}
E_K = \frac{1}{2}\int\!d^3\vx\, \rho\vu^2 \;,\qquad
E_Q = \frac{1}{8} \int\!d^3\vx\,\frac{\left(\grad_\vx\rho\right)^2}{\rho} \\
U = \int\!d^3\vx\,\Big[\rho h(\rho)-P(\rho)\Big]\;, \qquad
W = \frac{1}{2}\int\!d^3\vx\,\rho\Phi \;. \nonumber
\end{gather}
The pressure $P(\rho)$ is given by the equation of state
\begin{equation}
P= -\frac{1}{16} \rho^2 \;.
\end{equation}
Therefore, the internal energy reads
\begin{equation}
U = -\frac{1}{16} \int\!d^3\vx\,\rho^2 \;.
\end{equation}
We now look for configurations $(\rho,\vu)$ that minimize the Hamiltonian $H[\rho,\vu]$, that is, 
configurations which are stable, steady-state solutions of the GPP system. As we will discover, these configurations may exist only  if
some critical conditions are met and they crucially depend on the quartic coupling.

We follow  Ref. \cite{Chavanis:2011zi} and make a Gaussian ansatz for the density profile $\rho(t,\vx)$. 
We consider solutions with co-dimension $\overline D=1$ (planar symmetry or pancakes), $\overline D=2$ (cylindrical symmetry or
filaments) and $\overline D=3$ (spherical symmetry or halos), and use the corresponding cartesian, cylindrical and spherical coordinates. 
Therefore,
\begin{equation}
\rho(\vx,t) = C_{\overline D}(t) e^{-r^2/2R^2(t)} \;,
\end{equation}
where $R(t)$ is a time-dependent characteristic length and 
\begin{equation}
C_{\overline D}(t) = \frac{A_{\overline D}}{(2\pi)^{{\overline D}/2} R^{\overline D}} \;.
\end{equation}
Adopting a different profile, i.e. the solution of  Ref. \cite{ostriker:1964} for gas cylinders as a proxy for filaments for instance,
would only change the values of $\sigma_{\overline D}$, $\zeta_{\overline D}$, $\nu_{\overline D}$ by factors of order unity, but not affect scalings with
$A_{\overline D}$, $R$.
For pancakes ($\overline D=1$), $r=|\vr|$ becomes  the distance along the axis normal to the sheet plane whereas; for filaments ($\overline D=2$) $r$
is the radius in the two-dimensional plane  transverse to the filament; and for halos ($\overline D=3$) $r$ is the standard distance from  the halo
center.
Furthermore, $A_2\equiv \mathcal{M}$ and $A_1\equiv\Sigma$ are mass per unit length and surface density, respectively for  filaments and pancakes; 
$A_3\equiv M$ is a characteristic mass of the given halo.

Finally, we must also take the kinetic energy into account for the minimization of $H[\rho,\vu]$. Following \cite{Chavanis:2011zi} and using the ansatz
$\vu=(\dot{R}/R)\vr$ for the velocity field (which follows from a continuity argument) yields a kinetic energy $E_K\propto \frac{1}{2}\dot{R}^2$ regardless
of the exact shape of $\rho(\vx,t)$.
However, such an ansatz misses an important ingredient: dark matter particles generally moves along non-radial orbits
\cite{fillmore/goldreich:1984,ryden/gunn:1987,zaroubi/hoffman:1993,nusser:2001}.
Conservation of angular momentum prevents the object from collapsing down to a singularity. In practice, the kinetic energy associated to the solitonic
solutions should include -- at least for $\overline D=2$ and 3 -- a centrifugal potential which reflects the existence of non-radial motions and the
conservation of angular momentum:
\begin{equation}
E_K \propto \frac{1}{2} \left(\dot{R}^2+\frac{h^2}{R^2}\right) \;.
\end{equation}
For $\overline D=3$, $h=L$ is the magnitude of the angular momentum vector whereas, for $\overline D=2$, $h=L_z$ is the component along the filament
axis. For $\overline D=1$ however, symmetry implies that any net angular momentum would be directed towards the direction perpendicular to the pancake.
Hence, such a term would not prevent the pancake to collapse along the $\vr$-direction.
Nevertheless, since the exact magnitude of $h$ is unknown, we will ignore it in the subsequent analysis. This will not affect our result significantly
because it has the same scaling $\propto R^{-2}$ as the quantum pressure $E_Q$. However, we should bear in mind that it is generally present and, in the
case of standard CDM, leads to stable solutions for $\overline D=2$ and 3.

We emphasize that the symmetric solutions considered here describe axion solitonic cores inside halos, filaments or pancake. We will come back to
this point in \S~\ref{sec:lymanalpha} where we discuss implications for Lyman-$\alpha$ forest measurements.

\subsection{Halos}

We start with the spherical symmetric halos already studied thoroughly in Ref. \cite{Chavanis:2011zi}. 
Working with the rescaled coordinates and fields, we have (we omit again the tildes to avoid clutter)
\begin{eqnarray}
\rho=\frac{A_3}{(2\pi)^{3/2}R^3}e^{-r^2/2R^2}, ~~~r^2=x_1^2+x_2^2+x_3^2,\label{r3}
\end{eqnarray}
where $A_3=M_\text{c}/f^2$, and $M_\text{c}$ is the mass of the axions solitonic core, or ``axion star'', at the center of the halo.
The factor of $1/f^2$ arises because the rescaled density is $1/f^2$
times the physical density. In order to calculate the energy $E$ we need the potential $\Phi$, which satisfies
\begin{eqnarray}
\Delta\Phi=4\pi\tilde G \rho.  \label{f1}
\end{eqnarray}
Note that since 
\begin{eqnarray}
\rho\to A_3 \delta^{(3)}(\vr), ~~~~R\to 0,
\end{eqnarray}
we should impose the  condition on $\Phi$
\begin{eqnarray}
\Phi\to -\frac{A_3}{r}, ~~~~R\to 0.  \label{c1}
\end{eqnarray}
The solution to Eq.  (\ref{f1}) subject to the condition (\ref{c1}) is 
\begin{eqnarray}
\Phi=-\frac{\tilde G A_3}{r} {\rm erf}\left(\frac{r}{\sqrt{2}R}\right), \label{f3}
\end{eqnarray}
where $\rm{erf}(z)$ is the error function. 
Using, Eqs. (\ref{r3}) and (\ref{f3}), one finds 
\begin{eqnarray}
\bigg. E_Q&=&\sigma_3\frac{A_3}{R^2}\;, \qquad \sigma_3=\frac{3}{8}\;,\\
\bigg. U&=&\zeta_3\frac{A_3^2}{R^3}\;, \qquad ~\zeta_3=-\frac{1}{128 \pi^{3/2}}\;, \nonumber \\
\bigg. W&=&\nu_3 \frac{A_3^2}{R}\;,  \qquad ~\nu_3=-\frac{1}{2\sqrt{\pi}} \nonumber \;, 
\end{eqnarray}
so that the total potential $V=E_Q+U+W$ is 
\begin{eqnarray}
V(R)=\sigma_3\frac{A_3}{R^2}+\zeta_3\frac{A_3^2}{R^3}+\nu_3 \frac{\tilde G A_3^2}{R}.
\end{eqnarray}
Moving back to the dimensionful variables yields
\begin{align}
  V(R) &= \frac{\sigma_3}{f^2}\frac{M_\text{c}}{R^2} + \frac{\zeta_3}{f^4} \frac{M_\text{c}^2}{R^3}
  + \nu_3 \left(\frac{m}{f m_P}\right)^2 \frac{M_\text{c}^2}{R} \\
&= \frac{m^2}{f^2} \bigg(\frac{\sigma_3}{m^2} \frac{M_\text{c}}{R^2} + \frac{\zeta_3}{m^2 f^2} \frac{M_\text{c}^2}{R^3} 
+ \frac{\nu_3}{m_P^2}\frac{M_\text{c}^2}{R} \bigg) \nonumber \;.
\end{align}
In other words, this is equivalent to replacing $A_3$ by the physical mass $M$ and rescaling the dimensionless parameters 
$\sigma_3$, $\zeta_3$ and $\nu_3$ as follows:
\begin{align}
  \label{eq:scaling}
\sigma_3  &\to \frac{1}{m^2}\sigma_3 \; \\
\zeta_3 &\to \frac{1}{m^2 f^2}\zeta_3 \;, \nonumber \\
\nu_3 &\to \frac{1}{m_P^2} \nu_3 \nonumber \;.
\end{align}
The physical energy is obtained upon a multiplication by $f^2/m^2$, which would cancel the factor of $m^2/f^2$ in the second equality.
It is now very clear that the gravitational energy correctly behaves like $m_P^{-2}$ and, thus, vanishes in the limit $m_P\to \infty$. 
Conversely, the internal energy scales like $1/(m f)^2$ and, thus, vanishes when the decay constant tends towards $f\to\infty$
at fixed $m$.
Since the rescaling Eq. (\ref{eq:scaling}) is valid for any co-dimension $\overline D$, we shall hereafter express the potential $V$ directly
in term of the physical mass (or surface density etc.) and the dimensionless coupling $\sigma_{\overline D}$, $\zeta_{\overline D}$ and
$\nu_{\overline D}$.

Stable, steady solutions of the Hamiltonian are local minima of $V(R)$ and satisfy $\dot{R}=0$.
For $\overline D=3$, the critical radii for which $V'(R)=0$ turn out to be
\begin{equation}
R_\text{c,halo} = -\left(\frac{\sigma_3}{\nu_3 M_\text{c} }\right)\left(1 \pm\sqrt{1 -3 \frac{\zeta_3 \nu_3}{\sigma_3^2}M_\text{c}^2}\right) \;,
\end{equation}
Only for the solution with the minus sign in front of the square root

\begin{equation}
V''(R_\text{c,halo})=-\nu_3\frac{M^2_\text{c}}{R_\text{c,halo}^3}\left(1-\frac{3\zeta_3}{\nu_3 R_\text{c,halo}^2}\right)\;
\end{equation}
is positive and there is a stable solution for  masses $M_\text{c}\leq M_\text{c,max}$, 
\begin{equation}
M_\text{c,max} \simeq \frac{\sigma_3}{\sqrt{3 \zeta_3 \nu_3}} \;,
\end{equation}
that is (restoring dimensionful couplings)
\begin{align}
  M_\text{c,max} &= 7.1\times 10^{9}\, \frac{f_{17}}{m_{22}}\,\hmsun \\
  &= 7.1\times 10^{9}\, \frac{1}{\lambda_{96}}\,\hmsun \nonumber\;.
\end{align}
We observe that, in this case, gravity is essential to make the solitonic core stable at least  below some maximal mass.
Indeed, switching off gravity (i.e. $\nu_3=0$) one gets $V''(R_\text{c,halo})<0$ and halos are always unstable, as was already pointed out in
\cite{Chavanis:2011zi}.
By contrast, in the absence of self-interaction (i.e. $\zeta_3=0$), the quantum pressure always counteracts gravity, and the stability radius is
given by $R_\text{c,halo}=-2\sigma_3/\nu_3M_\text{c}$ at all mass.

\subsection{Pancakes}
\label{sec:pancakes}

\begin{figure*}
\includegraphics[width=0.49\textwidth]{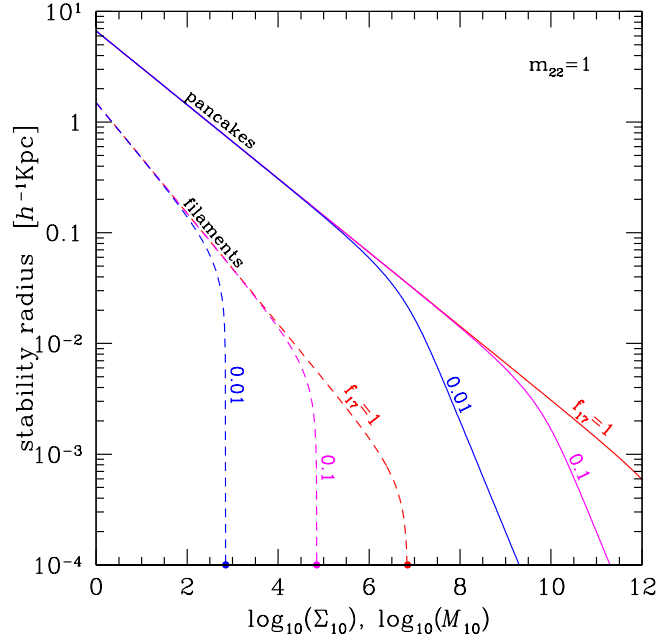}
\caption{Stability radius of the solitonic core for pancakes ($N=1$) and filaments ($N=2$) as a function of the core surface density $\Sigma_{10}$
  and mass per unit length $\mathcal{M}_{10}$, respectively. The solid and dashed curves show Eqs.(\ref{eq:Rpancake}) and (\ref{eq:Rfilament})
  for a decay constant $f_{17}=0.01$, 0.1 and 1. The points on the abscissa indicate the value $\mathcal{M}_{10}=\mathcal{M}_\text{c,max}$ for
  which the stability radius of the filament vanishes (see Eq. (\ref{eq:mmax})). An axion mass $m_{22}=1$ is assumed throughout for illustration.}
\label{fig:mradius}
\end{figure*}

For pancakes, we have
\begin{eqnarray}
\rho=\frac{A_1}{(2\pi)^{1/2} R}e^{-r^2/2R^2}, ~~~r^2=x_1^2,\label{rhopancake}
\end{eqnarray}
where $A_1=\Sigma_\text{c}/f^2$ is the  surface density of the solitonic core inside the axion pancake.
Now we have, 
\begin{eqnarray}
\rho\to A_1\delta(x_1), ~~~~R\to 0,
\end{eqnarray}
and therefore we should  impose the condition for $\Phi$
\begin{eqnarray}
\Phi\to 2\pi  A_1 |x_1|, ~~~~R\to 0.  \label{c3}
\end{eqnarray}
The solution to Eq. (\ref{f1}) which satisfies (\ref{c3}) is given by
\begin{eqnarray}
\Phi=4\pi A_1\left\{\frac{R}{\sqrt{2\pi}}e^{-r^2/2R^2}+\frac{r}{2}\,{\rm erf}\left(\frac{r}{\sqrt{2}\, R}\right)\right\}.
\end{eqnarray}
Integrating over the range $-\infty < r < +\infty$, we obtain
\begin{align}
\bigg. E_Q&=\sigma_1\frac{\Sigma_\text{c}}{R^2}\;, \qquad\quad\sigma_1=\frac{1}{8} \;, \\ 
\bigg. U&=\zeta_1\frac{\Sigma_\text{c}^2}{R}\;, \qquad\quad ~\zeta_1=-\frac{1}{32\sqrt{\pi}} \nonumber \;, \\ 
\bigg. W&=\nu_1 \Sigma_\text{c}^2 \, R\;, \qquad ~~\nu_1=2\sqrt{\pi} \nonumber \;, 
\end{align}
and the potential $V=E_Q+U+W$ is given by 
\begin{equation}
V(R)=\sigma_1\frac{\Sigma_\text{c}}{R^2}+\zeta_1\frac{\Sigma_\text{c}^2}{R}+\nu_1\Sigma_\text{c}^2\, R \;.
\end{equation}
Axion filamentary cores collapse and stabilize at the critical radius
\begin{align}
  \label{eq:Rpancake}
R_\text{c,pancake} &= \sqrt[3]{\frac{\sigma_1}{\nu_1\Sigma_\text{c}}}\bigg[-\alpha \left(1+\sqrt{1 +\alpha^3}\right)^{-1/3} \\
&\quad + \left(1+\sqrt{1 +\alpha^3}\right)^{1/3}\bigg] \nonumber\;,
\end{align}
where the (positive definite) dimensionless ratio $\alpha$ is
\begin{equation}
  \label{eq:alpha}
\alpha \equiv \frac{|\zeta_1|}{3}\sqrt[3]{\frac{\Sigma_\text{c}^2}{\nu_1\sigma_1^2}} \;.
\end{equation}
One can check that
\begin{equation}
V''\left(R_\text{c,pancake}\right)>0
\end{equation}
regardless of the value of $m$, $f$ or $\Sigma_\text{c}$.
Introducing a normalized surface density
\begin{equation}
\Sigma_{10}\equiv \frac{\Sigma_\text{c}}{10^{10}\surfdens} \;,
\end{equation}
the dimensionless ratio becomes
\begin{align}
\alpha &= 1.6\times 10^{-2}\frac{|\lambda|}{m^2}\left(\frac{m_P}{m}\right)^{2/3}\Sigma_\text{c}^{2/3} \\
&= 1.8\times 10^{-9}\, \frac{\lambda_{96}\,\Sigma_{10}^{2/3}}{m_{22}^{8/3}} \nonumber \;.
\end{align}
The one-dimensional mass-radius depends sensitively on the value of $\alpha$.
So long as $\alpha$ is small, $R_\text{c,pancake}$ is approximately given by
\begin{equation}
\label{eq:RSlowlambda}
R_\text{c,pancake}\approx 6.7\, m_{22}^{-2/3}\, \Sigma_{10}^{-1/3} \hkpc
\end{equation}
independently of the quartic coupling $\lambda$.
This reflects the fact that, for small values of $\lambda$ and/or $\Sigma_\text{c}$, gravity is balanced by 
the quantum pressure and the ultra-light  self-interaction does not play any role.
The latter becomes important when $\alpha\gtrsim 1$ or, equivalently, when
\begin{equation}
\Sigma_{10} \gtrsim 1.4\times 10^{13}\, \frac{m_{22}^4}{\lambda_{96}^{3/2}}\;.
\end{equation}
For our fiducial choice of $m$ and $f$, this occurs when the surface density exceeds $\sim 10^{23}\surfdens$.
In this regime, the one-dimensional  stability radius is given by
\begin{equation}
\label{eq:RShighlambda}
R_\text{c,pancake} \approx 2.0\times 10^9\, \frac{m_{22}^2}{\lambda_{96}\,\Sigma_{10}}\hkpc \;.
\end{equation}
Therefore, $R\to 0$ in the limit $\lambda\to \infty$ as expected.
This double power-law behavior is clearly seen in Fig.\ref{fig:mradius}, where the
stability radius of pancakes, Eq. (\ref{eq:Rpancake}), is shown for different values of the decay constant.

What about pancakes immersed in higher dimensions? We have seen that, in the absence of gravity, they are unstable against transverse fluctuations.
Let us scrutinize the fate of these pancakes when gravity is included.
For this purpose, inspired by Ref.  \cite{2017PhRvL.118x4101K}, we study the transverse instability modes of lower-dimensional objects embedded in a
higher-dimensional space as these modes can lead to the breakup of the solitons through the so-called Landau dynamics approach  \cite{K1,K2}.
In practice, this amounts to studying the semi-classical dynamics of the solitary wave as a quasi-particle.
The starting point is the one-dimensional equation

\begin{equation} 
\label{el}
i \partial_t\psi+ \Delta_{x}\psi +  |\psi|^{2}\psi-\Phi(x)\psi=0\;, 
\end{equation}
where $\Phi(x)$ is an external potential, which we will later identify with the gravitational potential.
In the limit of vanishing gravity, the  pancake-like solitonic solution reads

\begin{equation}
\label{a1}
\psi(x,t)= \frac{\sqrt{2\omega+U^2/2}\,\, e^{i\left(Ux/2+\omega t\right)}}{\cosh\!\left(\sqrt{\omega+U^2/4}\, (x-U t-x_0)\right)} \;,
\end{equation} 
where $x_0$ is the center of the solitonic pancake and $U=\partial_t {x}_0$ is its constant velocity.
The energy associated with this system is $\Omega_1=H_1+\omega N$, that is,

\begin{equation}
\label{a2}
\Omega_{1} =\int_{-\infty}^\infty\! dx\,\bigg[\left|\partial_x \psi\right|^2
-\frac{1}{2}\left(\left|\psi\right|^4-2\omega \left|\psi\right|^2\right)\bigg]\;. 
\end{equation}
Inserting Eq. (\ref{a1}) into (\ref{a2}) one finds

\begin{equation}
\Omega_1= \frac{1}{3}\Big(4\omega+U^2\Big)^{3/2}\;. 
\end{equation}
Let us now assume that the solution (\ref{a1}) is immersed in a two-dimensional ambient space with coordinates $(x,y)$.
In this case, the energy reads $\Omega_2=H_2+\omega N$, i.e.

\begin{align}
\label{a3}
\Omega_{2} & =\int_{-\infty}^\infty\! dx\int_{-\infty}^\infty\! dy\,\bigg[\left|\partial_x \psi\right|^2+\left|\partial_y \psi\right|^2 \\
  &\quad -\frac{1}{2}\left(\left|\psi\right|^4-2\omega \left|\psi\right|^2\right)\bigg] \nonumber\;. 
\end{align}
On assuming  that  $x_0=x_0(t,y)$, the energy (\ref{a3}) simplifies to

\begin{equation}
\Omega_2=\int_{-\infty}^\infty\! dy\, \frac{2+(\partial_y x_0)^2}{6}\Big\{4\omega+(\partial_t x_0)^2\Big\}^{3/2}\;. 
\end{equation}
Therefore, if both the Hamiltonian $H_2$ and the number of particle $N$ are conserved, $\Omega_2$ is also conserved and we obtain

\begin{align}
  0 &=\frac{d \Omega_2}{dt} \\
  &=\int_{-\infty}^\infty\! dy\,
\frac{\partial}{\partial t}\left\{\frac{2+(\partial_y x_0)^2}{6}\Big(4\omega+(\partial_t x_0)^2\Big)^{3/2}\right\} \nonumber\;.
\end{align}
This leads to the equation
\begin{multline}
6 \partial_t^2 x_0-6 \partial_y x_0\partial_t x_0\partial_t\partial_y x_0 \\
-8 \omega \partial^2_y x_0-2 (\partial_t x_0)^2\partial^2_{y}x_0+3\partial_t^2 x_0(\partial_y x_0)^2=0\;. 
\end{multline}
At the linearized level,  $x_0$ satisfies the equation
\begin{eqnarray}
\partial^2_t x_0-\frac{4\omega}{3} \partial_y^2 x_0=0\;.
\end{eqnarray}
Clearly, for $\omega<0$ (and one can always reduce oneself to this case by a Galilean transformation), this equation is of elliptic-type
and leads to instabilities.
This indeed confirms the analysis of \S~\ref{sec:nograv_stability}.

In the presence of a gravitational  potential $\Phi(x)$, the Landau dynamics approach  assumes that the energy $\Omega$ is an
adiabatic invariant, that is, $\Phi(x)$ varies slowly and (minus) the chemical potential $\omega$ is replaced by $\omega+\Phi(x)$. 
In this case, the corresponding energy is 
\begin{align}
\label{a4}
\Omega_{2} &=\int_{-\infty}^\infty\! dx\int_{-\infty}^\infty\! dy\,\bigg[\left|\partial_x \psi\right|^2+\left|\partial_y \psi\right|^2 \\
  &\qquad -\frac{1}{2}\left(\left|\psi\right|^4-2(\omega+\Phi(x)) \left|\psi\right|^2\right)\bigg] \nonumber 
\end{align}
and its variation gives rise to Eq. (\ref{el}). For the solution (\ref{a1}), we find  
\begin{eqnarray}
\Omega_2&=&\int_{-\infty}^\infty\! dy \, \frac{2+(\partial_y x_0)^2}{6}\Big(4\omega+4\Phi(x)+(\partial_t x_0)^2\Big)^{3/2}\; .
\nonumber\\
&& \label{H2}
\end{eqnarray}
The Landau dynamics approach then assumes that 
\begin{align}
0 & =\int_{-\infty}^\infty\! dy\, 
\frac{\partial}{\partial t}\!\bigg\{\frac{2+(\partial_y x_0)^2}{6} \\
& \qquad \times \Big(4\omega+4\Phi(x)+(\partial_t x_0)^2\Big)^{3/2}\bigg\} \nonumber \;,
\end{align}
leading to the equation
\begin{multline}
6 \partial_t^2 x_0-6 \partial_y x_0\partial_t x_0\partial_t\partial_y x_0 -6 \left(2-
\partial_y x_0^2\right)\frac{\partial \Phi}{\partial x_0}\\
-8 \omega \partial^2_y x_0-2 (\partial_t x_0)^2\partial^2_{y}x_0+3\partial_t^2 x_0(\partial_y x_0)^2=0\;. 
\end{multline}
At the linear level (assuming that $V$ is of the same order as $x_0^2$), we get 
that $x_0$ satisfies 
\begin{eqnarray}
\partial^2_t x_0-\frac{4\omega}{3} \partial_y^2 x_0=2\frac{\partial \Phi}{\partial x_0}\,. \label{vx}
\end{eqnarray}
In the thin wall approximation, the potential of a sharply localized  pancake along the direction $x$ will have  a gravitational potential
$\Phi\sim |x|$ (see Eq.~(\ref{c3})). Therefore, Eq. (\ref{vx}) is a wave equation with a constant external source.
Hence, there will be again instability as in the case with no gravity.

We should stress that, in the above discussion, we have taken the gravitational potential to be the potential resulting from the Poisson
equation with a source given by the pancake-like soliton Eq. (\ref{a1}).
Therefore, the dependence of this solution on the transverse direction $y$ appears through the central position $x_0(t,y)$ of the pancake.
However, the gravitational potential is in principle sourced by the fluctuations of the soliton itself.
Our analysis above assumes that the energy functional (\ref{a4}) is still an adiabatic invariant even if $\Phi$ is sourced by $\psi$.
Within this approximation, we conclude that gravity is not able to halt the transverse instabilities of the pancake-like profiles.
It would be interesting to investigate what is the role of gravity when the gravitational potential is fully and consistently taken account
since the corresponding non-locality might help the stabilization \cite{greek}.

\subsection{Filaments}
\label{sec:filaments}

Filaments have $N=2$, so that their density profile reads
\begin{eqnarray}
  \rho=\frac{A_2}{(2\pi) R^2}e^{-r^2/2R^2}, ~~~r^2=x_1^2+x_2^2 \;,
  \label{rhofilament}
\end{eqnarray}
where $A_2=\mathcal{M}_\text{c}/f^2$ is the  mass per unit length of the solitonic core at the center of the filament.
Again since, 
\begin{eqnarray}
\rho\to A_2\delta^{(2)}(\vr), ~~~~R\to 0,
\end{eqnarray}
we should impose the condition for $\Phi$
\begin{eqnarray}
\Phi\to 2 A_2 \ln \, r, ~~~~R\to 0.  \label{c2}
\end{eqnarray}
 The solution to Eq. (\ref{f1}) which satisfies (\ref{c2}) is
\begin{eqnarray}
\Phi=-A_2 {\rm Ei}\left(\frac{r^2}{2R^2}\right)+2A_2\,  \ln\, r,
\end{eqnarray}
where ${\rm Ei}(z)=-\int_{-z}^\infty dt \, e^{-t}/t$ is the exponential integral function.  
Discarding a (irrelevant) constant contribution to the gravitational energy, we find
\begin{align}
\bigg. E_Q &=\sigma_2\frac{{\mathcal{M}_\text{c}}}{R^2}\;, \qquad\qquad\sigma_2=\frac{1}{4} \;, \\
\bigg. U &=\zeta_2\frac{\mathcal{M}_\text{c}^2}{R^2}\;, \qquad\qquad \zeta_2=-\frac{1}{64 \pi}, \nonumber \\
\bigg. W &=\nu_2 \mathcal{M}_\text{c}^2 \, \ln \, R\;, \qquad ~\nu_2=1 \nonumber \;,
\end{align}
so that the potential $V$ becomes 
\begin{equation}
V(R)=\sigma_2\frac{\mathcal{M}_\text{c}}{R^2}+\zeta_2\frac{\mathcal{M}_\text{c}^2}{R^2}+\nu_2\mathcal{M}_\text{c}^2\,\ln\, R\;.
\end{equation}
For standard CDM, only the last term is present but, owing to the centrifugal barrier, the Hamiltonian would
exhibit stable filamentary solutions for any value of the mass per unit length $\mathcal{M}$.

This should be contrasted to the axion case, for which there exist stable filamentary configurations only
below a critical mass per unit length.
Namely, introducing the dimensionless quantity
\begin{equation}
\mathcal{M}_{10}\equiv \frac{\mathcal{M}_\text{c}}{10^{10}\masslength} \;,
\end{equation}
the critical radius is given by
\begin{align}
    \label{eq:Rfilament}
  R_\text{c,filament} &= \sqrt{\frac{2}{\nu_2 \mathcal{M}_\text{c}}\left(\sigma_2+\zeta_2\mathcal{M}_\text{c}\right)} \\
  &= 1.5 \Big(m_{22}^2\mathcal{M}_{10}\Big)^{-1/2} \nonumber \\
  &\qquad \times \sqrt{1- 1.4\times 10^{-7} \frac{\mathcal{M}_{10}}{f_{17}^2}} \hkpc \nonumber \;.
\end{align}
and exists only if $\sigma_2+\zeta_2\mathcal{M}_\text{c}$ is positive.
This yields the condition $\mathcal{M}_{10}\leq \mathcal{M}_\text{c,max}$, where the critical mass per unit length
(in unit of $10^{10}\masslength$) is
\begin{equation}
  \label{eq:mmax}
\mathcal{M}_\text{c,max} =7.0\times 10^6\,f_{17}^2 = 7.0\times 10^6 \frac{m_{22}^2}{\lambda_{96}} \;.
\end{equation}
A straightforward calculation shows that
\begin{equation}
V''\!\left(R_\text{c,filament}\right)>0 \;,
\end{equation}
so that the solution is stable. Namely, the quantum pressure overcomes both gravity and the attractive self-interaction, 
and the filament does not collapse. By contrast, for a mass per unit length $\mathcal{M}_{10}\gtrsim \mathcal{M}_\text{c,max}$,
the filaments are unstable owing to the self-interaction being stronger than the quantum pressure.
This is in agreement with the results of the previous section where it was shown that above a critical number filaments are
unstable in the absence of gravity. For illustration, Eq. (\ref{eq:Rfilament}) is shown in Fig.\ref{fig:mradius} as a function
of $\mathcal{M}_{10}$ for a few values of the decay constant $f_{17}$.
When $\zeta_2\mathcal{M}_\text{c}\ll \sigma_2$, the stability radius is given by
\begin{equation}
\label{eq:Rfilament1}
R_\text{c,filament}\approx \sqrt{\frac{2\sigma_2}{\nu_2\mathcal{M}_\text{c}}}
\end{equation}
independently of the strength $\zeta_2$ of the self-interaction. For a decay constant $f_{17}\gtrsim 0.01$, this is a very good
approximation so long as $\mathcal{M}_{10}\lesssim 2$.

On which timescale do these filamentary cores collapse if $\sigma_2+\zeta_2\mathcal{M}_\text{c}$ is negative ?
Upon making the ansatz $\vu=(\dot{R}/R)\vr$  \cite{Chavanis:2011zi} and substituting into the expression of $E_K$, the 
Lagrangian reads
\begin{align}
  L(R,\dot{R},t) &\approx E_K-E_Q-U-W \\
  &=\mathcal{M}_\text{c} \dot{R}^2-V(R) \nonumber \\
  &= \mathcal{M}_\text{c} \dot{R}^2 - \frac{\sigma_2\mathcal{M}_\text{c}+\zeta_2\mathcal{M}_\text{c}^2}{R^2}-\nu_2\mathcal{M}_\text{c}^2\ln R \nonumber\;.
\end{align}
The corresponding equation of motion is

\begin{equation}
\mathcal{M}_\text{c}\ddot{R}=-\frac{1}{2}V'(R)\;,
\end{equation}
whose solution is

\begin{equation}
\mathcal{M}_\text{c}^{-1/2}(t-t_i)=\int_{R(t)}^{R_i}\frac{dR}{\left(V(R_i)-V(R)\right)^{1/2}}\;.
\end{equation}
Here, $R_i$ is the radius of the filament at initial time $t_i$ and we have assumed an initial velocity $\dot{R}_i\equiv 0$.
The collapse time $t_\text{coll}$ is obtained from the requirement $R(t_\text{coll})=0$, so that the solution can be re-expressed as 

\begin{equation}
\mathcal{M}_\text{c}^{-1/2}(t_{\rm coll}-t_i)=\int_{0}^{R_i}\frac{dR}{\left(V(R_i)-V(R)\right)^{1/2}}\;.
\end{equation}
Ignoring the self-gravity contribution to the potential (which scales only logarithmic), the radius behaves like 

\begin{equation}
R(t)\simeq \sqrt{2}\left|\sigma_2+\zeta_2\mathcal{M}_\text{c}\right|^{1/4}(t_{\rm coll}-t)^{1/2}
\end{equation}
close to the collapse time. Taking $t_i\approx 0$, we obtain the timescale $t_\text{coll}$ over which a filament of initial thickness
$R_i$ disappears. Comparing this timescale to the age of the Universe, $t_0\approx 1.4\times 10^{10}$ yr, and restoring dimensionful
mass and couplings, we obtain
\begin{align}
  \frac{t_\text{coll}}{t_0} &\sim \frac{m R_i^2}{2t_0\left|\sigma_2+\zeta_2\mathcal{M}_\text{c}/f^2\right|^{1/2}} \\
    &\simeq 4.8\,\frac{m_{22}}{\big(\mathcal{M}_\text{c}/f_{17}^2\big)^{1/2}}\left(\frac{R_i}{\hkpc}\right)^2 \nonumber \;.
\end{align}
Our naive estimate agrees with the scaling $t_\text{coll}\propto M_\text{c}^{-1/2}$ found by \cite{chavanis:2016} in the limit
$M\gg M_\text{max}$.
Note that we have neglected the expansion of the universe and, moreover, assumed that the mass per unit length is conserved.
Filaments with an axion core of mass $\mathcal{M}_\text{c}\gtrsim \mathcal{M}_\text{c,max}$ collapse on a timescale $t_\text{coll}<t_0$
so long as their initial (proper) radius is $R_i\lesssim 10\hkpc$.
As we shall see in \S\ref{sec:lymanalpha} however, such filaments are so rare when $f_{17}\gtrsim 0.01$ than this instability can be
safely ignored.

\section{Signature in the Lyman-$\alpha$ forest}
\label{sec:lymanalpha}

We will now investigate whether the instabilities found above can leave a detectable signature in the Lyman-$\alpha$ forest.
For sake of illustration, we shall focus on the $D=2$ solutions with cylindrical symmetry, as a proxy to the dark matter filaments
traced by the Lyman-$\alpha$ forest.

\subsection{Axion core -- filament mass relation}

In order to relate the solitonic, filamentary solution discussed in \S~\ref{sec:filaments} to a high-redshift filament of gas seen
in Lyman-$\alpha$ absorption, we must take into account the fact that the axion solitons are surrounded by a haze of virialized
axions which extend much farther than the solitonic core.
The latter is dubbed ``axion star'' in the literature when it refers to the axion core at the center of spherical halos. Numerical
simulations of the GPP system have established that the core-halo mass relation is given by $M_c\propto M_\text{halo}^{1/3}$
\cite{viel10,schive/liao/etal:2014,schwabe/niemeyer/etal:2016,veltmaat/niemeyer:2016,du/behrens/etal:2017}.
We will assume that the same relation holds for filamentary configurations but, since it has thus far not been measured from
simulations, we will treat the overall normalization scale as a free parameter. Namely, we write
\begin{equation}
 \label{eq:core-mass}
\mathcal{M}_\text{c} = 1.1\times 10^{-3}A_\text{c}\left(\frac{\mathcal{M}_g}{4.4\times 10^{-3}f_g A_\text{c}}\right)^{n_\text{c}}\;.
\end{equation}
where the baryonic or gas mass $\mathcal{M}_g$ per unit length and the core mass $\mathcal{M}_c$ are both in unit of $10^{10}\masslength$,
$f_g\sim 0.2$ is the baryon mass fraction, $n_\text{c}$  is the powerlaw index and $A_\text{c}$ is an overall normalization factor
(which may generally depend on $f_{17}$). Simulations suggest that $A_\text{c}=1$ and $n_\text{c}=1/3$ for non-interacting axion
cores within dark matter halos.

Filaments with a baryonic mass per unit length $\mathcal{M}\gtrsim 10^{12}\masslength$ are very rare at redshift $z\sim 3$ as shown,
e.g., by the excursion set analysis of \cite{Shen:2005wd}.
Therefore, if the above axion core-filament mass relation holds with $A_\text{c}=1$, it is unlikely that a filamentary
object with a axion core mass $\mathcal{M}_\text{c}\gtrsim 0.1$ is observed as an absorption feature in the Lyman-$\alpha$ forest.
Such a mass would still be $\sim 4$ order of magnitude below the critical mass $\mathcal{M}_\text{c,max}$ for a decay constant as
low as $f_{17}=0.01$. Therefore, axion self-interactions will not imprint any signature in the Lyman-$\alpha$ forest unless the
decay constant is $f_{17}\ll 0.01$. 

Notwithstanding, since the core - mass relation (\ref{eq:core-mass}) is uncertain for filaments, it is interesting to explore the
impact of the solitonic axion core on Lyman-$\alpha$ absorption features as a function of the normalization $A_\text{c}$.
In principle, we should also consider possible variations in $n_\text{c}$ (the analysis of \cite{du/behrens/etal:2017} suggest that
$n_\text{c}\approx 0.4$ is a better fit to the numerical data). For simplicity however, we shall hereafter assume a unique value
$n_\text{c}=1/3$ of the powerlaw index.

\subsection{Hydrostatic approximation}

Absorption features in the Lyman-$\alpha$ forest are characterized by their column density $N_\text{\tiny HI}$ of neutral 
hydrogen {\small HI} (and their width, but we shall ignore it here).
While it is straightforward to estimate a column density for a spherically symmetric absorber \cite[see][for instance]{Schaye:2001me},
the task is more challenging in $D=2$ because both the pressure gradient $\grad_\vr P \sim -c_s^2 \rho\rvh/r$ and the gravitational
acceleration ${\bf g}\sim  -G\mathcal{M}\rvh/r$ scale like $1/r$.

To estimate the characteristic size of the Lyman-$\alpha$ absorption feature and, thereby, assign a column density $N_\text{\tiny HI}$ 
to dark matter filaments, we assume that the gas is in hydrostatic equilibrium so that the characteristic size of the feature changes 
only slowly with time. This is a reasonable approximation so long as the sound-crossing and free-fall timescales are short compared to
the age of the Universe, which is the case for $N_\text{\tiny HI}\gtrsim 10^{14}\cmm$.

Let $\rho_g$, $T_g$ and $P_g$ be the gas density, temperature and pressure, $\rho_c$ the density of dark matter and $\Phi$ the total
gravitational potential.
Hydrodynamical simulations have shown that, in the low density, highly ionized IGM traced by the Lyman-$\alpha$ forest
($\delta\rho_g/\bar\rho_g\lesssim 5$), equilibrium between photoionization and adiabatic cooling leads to a tight power-law relation
between the gas temperature and density of the form $T_g=\hat T_g (\rho_g/\bar\rho_g)^{\gamma-1}$, where $\hat T_g$ is the gas temperature
at mean gas density $\bar\rho_g$ and the value of the exponent $\gamma$ depends on the details of the process \cite{Hui:1997dp}. 
Using the ideal gas law, the low density is thus described by a polytropic equation of state $P_g = K \rho^\gamma$, with 
$K = \frac{1}{\mu m_p} \hat T_g\bar\rho_g^{1-\gamma}$. Here, $m_p$ is the proton mass and $\mu\approx 0.6$ is the mean molecular
mass appropriate to a fully ionized plasma of primordial abundance.

The polytropic equation of state, together with the hydrostatic equilibrium
\begin{equation}
\frac{1}{\rho_g}\grad_\vr P_g = - \grad_\vr\Phi \;,
\end{equation}
the continuity and the Poisson equation, eventually yields the cylindrical Lane-Emden equation with the dark matter component as 
an external source:
\begin{equation}
\label{eq:lanemden}
\frac{1}{\xi} \frac{d}{d\xi}\left(\xi\frac{d\theta}{d\xi}\right)+\theta^n = -\frac{\rho_\text{DM}}{\rho_0} \;,
\end{equation}
where the polytropic index satisfies $n=1/(\gamma-1)$ and $\rho_0$ is the gas density on the symmetry axis ($r=0$). 
The radial coordinate $r$ transverse to the filament has been rescaled according to $r=\alpha_n \xi$, where \cite{ostriker:1964}
\begin{equation}
\alpha_n^2 = \frac{K(n+1) \rho_0^{1/n-1}}{4\pi G} \;.
\end{equation}
Furthermore, the gas density is written in terms of the dimensionless function $\theta(\xi)$ such that $\rho_g(\xi)\equiv \rho_0 \theta(\xi)^n$.

Solutions to the cylindrical Lane-Emden equation are subject to the initial condtiions $\theta(0)=1$ and $\theta'(0)=0$. 
In the absence of a source, they can be found numerically for arbitrary values of the polytropic index $n$ as shown in \cite{ostriker:1964}.
The abscissa $\xi_1$ of the first zero
of $\theta$, i.e. $\theta(\xi_1)=0$, corresponds to $P_g(\xi_1)=\rho_g(\xi_1)=0$ and, therefore, defines the size of the gas filament. The size of
the filament strongly depends on $n$. Ignoring the source term, one finds $\xi_1\approx 2.405$ for $n=1$ ($\gamma=2$), $\xi_1\approx 2.92$
for $n=2$ ($\gamma=1.5$) whereas, in the limit $n\to\infty$ (corresponding to $\gamma\to 1$), $\xi_1$ diverges \cite{ostriker:1964}.

\begin{figure*}
\includegraphics[width=0.49\textwidth]{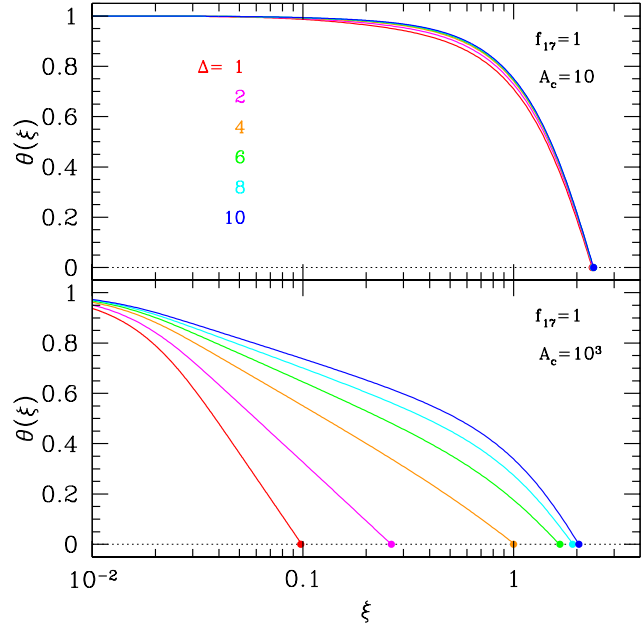}
\caption{Solutions to the Lane-Emden equation Eq. (\ref{eq:lanemden}) as a function of the gas overdensity $\Delta$ at the center of the
  filament, which takes values in the range $1\leq\Delta\leq 10$.
  The top and bottom panel assume a normalization $A_\text{c}=10$ and $10^3$ for the axion core - filament mass, respectively.
  They display the normalized gas profile $\theta(\xi)$ as a function of the dimensionless radius $\xi$.
  The points indicate the abscissa $\xi_1$ of the first zero-crossing, which should be interpreted as the radius of the filament.
  The gas equation of state is a polytrope with index $n=1$ (i.e. an exponent $\gamma=2$).
  The axion profile is the solitonic solution Eq. (\ref{rhofilament}), with a radius $R=R_\text{c,filament}$ given by the stability condition
  Eq.~(\ref{eq:Rfilament}).
  An axion mass $m_{22}=1$ and a decay constant $f_{17}=1$ are assumed for the calculation of $R_\text{c,filament}$.
  For the range of axion core masses obtained here (see text), the axion self-interaction is very weak so that $R_\text{c,filament}$ is well
  approximated by Eq.~(\ref{eq:Rfilament1}).}
\label{fig:lanemden}
\end{figure*}

We will now outline how we assign a {\small HI} column density to a filament, and provide some estimate for their abundance, before we solve
the inhomogeneous Lane-Emden equation (\ref{eq:lanemden}).

\subsection{Column densities and abundances}

Firstly, we use results from hydrodynamical simulations and write the number density of neutral hydrogen in the IGM as
$n_\text{\tiny HI}=\hat n_\text{\tiny HI}(\rho_g/\bar\rho_g)^\beta$, where \cite{Hui:1996fh}
\begin{align}
  \label{eq:avnhi}
  \hat n_\text{\tiny HI} &= 7.0\times 10^{-11}\left(\frac{\Gamma_\text{\tiny phot}}{10^{-12}\ {\rm s^{-1}}}\right)^{-1}
  \left(\frac{\hat T_g}{10^4\Kel}\right)^{-0.7} \\
  & \qquad \times 
  \left(\frac{\Omega_b h^2}{0.0227}\right)^2\left(\frac{1+z}{4}\right)^6\cmmm \nonumber 
\end{align}
is the {\small HI} number density at mean gas density $\bar\rho_g$.
Here, $\beta=2-0.7(\gamma-1)$ and $\Gamma_\text{\tiny phot}$ is the photoionization rate \cite{Hui:1996fh}.
The numerics assume a primordial helium abundance of $Y=0.248$.
The exact values of the temperature $\hat T_g$ at mean gas density, the exponent $\gamma$ and the photoionization rate $\Gamma_\text{\tiny phot}$
at a given redshift depend on the details of the reionization history and the cosmology.
In the redshift range $z\sim 2 - 4$, observations of high-redshift quasars indicate that
$\hat T_g\sim 10^4\Kel$, $\Gamma_\text{\tiny phot}\sim 10^{-11} - 10^{-12}\ {\rm s^{-1}}$ and $\gamma$ is in the range 1 - 1.6
(see \cite{viel19} for a recent review on the properties of the high-redshift IGM).
We shall adopt the fiducial values $\hat T_g=10^4\Kel$, $\Gamma_\text{\tiny phot}= 10^{-12}\ {\rm s^{-1}}$,
$\Omega_bh^2=0.0227$ and $\gamma=2$ (i.e. $n=1$).
Arguably, $\gamma=2$ is somewhat in tension with observations but, as we shall see below, this choice will enable us to obtain an analytic
expression for the Green function of the cylindrical Lane-Emden equation.

For these fiducial values, the normalization $\alpha_n^2\propto \bar\rho_g^{-1} \Delta^{1/n-1}$, where $\Delta=\rho_0/\bar\rho_g$ is the ratio of
central to average gas density, is given by
\begin{align}
  \label{eq:alpha1}
  \alpha_1 &= 1.8\times 10^{37} \left(\frac{1+z}{4}\right)^{-3/2}\, {\rm GeV}^{-1} \\
  &= 78 \left(\frac{1+z}{4}\right)^{-3/2}\hkpc \nonumber
\end{align}
and does not depend on $\Delta$. Using Eq.(\ref{eq:lanemden}), the mass of gas $\mathcal{M}_g$ per unit length reads \cite{ostriker:1964}
\begin{align}
\label{eq:Mg}
\mathcal{M}_g &= 2\pi \int_0^{r_1}\! dr\,r\rho_g(r) \\
&= \Big. 2\pi \rho_0\alpha_1^2 \big\lvert \xi_1 \theta'(\xi_1)\big\lvert \nonumber \\
&= \Big. 3.2\times 10^{10}\,\Delta\, \big\lvert \xi_1 \theta'(\xi_1)\big\lvert \masslength \nonumber \;,
\end{align}
where $r_1 = \alpha_1 \xi_1$. For $n=1$, the filament baryonic mass $\mathcal{M}_g$ grows linearly with the central gas density.
Note also that it is always independent of redshift.

Assuming that the line-of sight to the distant quasar is perpendicular to the symmetry axis and goes through the origin, the {\small HI} column
density of the filament is 
\begin{align}
  \label{eq:NHI}
N_\text{\tiny HI} &= 2 \int_0^{r_1}\!dr\,n_\text{\tiny HI}(r,\theta,z) \\
& = 2 \alpha_n \hat n_\text{\tiny HI} \Delta^\beta\left(\frac{1+z}{4}\right)^{9/2} \int_0^{\xi_1}\!d\xi\,\theta(\xi)^{n\beta} \nonumber \\
&= 4.87\times 10^{13} \Delta^2\left(\frac{1+z}{4}\right)^{9/2} \int_0^{\xi_1}\!d\xi\,\theta(\xi)^2\,\cmm  \nonumber
\end{align}
where, in the last equality, we have specialized the result to our fiducial choice of parameters, for which $\beta=2$.
The homogeneous Lane-Emden equation gives $|\xi_1\theta'(\xi_1)|\simeq 2.25$ and $\int\!\! d\xi\, \theta(\xi)^2\simeq 1.35$.
For a filament with overdensity $\Delta=2$ at redshift $z=3$, this yields
\begin{align}
  \mathcal{M}_g &\simeq 1.4\times 10^{11}\masslength \\
  N_\text{\tiny HI} &\simeq 2.6\times 10^{14}\cmm \nonumber \\
  \alpha_1\xi_1 &\simeq 188 \hkpc \nonumber \;,
\end{align}
where $\alpha_1\xi_1$ is the proper radius of the filament.
These values are consistent with the typical baryonic mass, column density and width of filaments identified in hydrodynamical
simulations of the low density, high-redshift IGM \cite{petitjean/muket/etal:1995,escude/cen/etal:1996}.

Of course, most of the sight lines do not actually pass through the center and perpendicularly to the filament.
Therefore, our estimate of $N_\text{\tiny HI}$ is only indicative as one shall expect a wide range of {\small HI} column densities for
each filament depending on the impact parameter etc. of the sight lines.
In fact, the column density distribution (per unit absorption length) $f(N_\text{\tiny HI},z)$, which is approximately given by
\cite{erkal/gnedin/etal:2012}
\begin{equation}
  \label{eq:fHIz}
  f(N_\text{\tiny HI},z) =\frac{1}{H_0} \int\!\!d\ln M\, n(M,z)\, \frac{d\sigma}{dN_\text{\tiny HI}}\!(M,N_\text{\tiny HI},z)
\end{equation}
where $d\sigma/dN_\text{\tiny HI}\propto R_\text{filament}^2$ is the differential cross-section for producing absorbers with column density
$N_\text{\tiny HI}$, should reflect the gas density profile around a filament (see, e.g., \cite{milgrom:1988}).
This motivates the calculation of gas density profiles presented in \S\ref{sec:solvingLE}.

To estimate the abundance of such absorption lines, we approximate filaments as objets that have virialized along two dimensions (rather
than three as for halos).
In the excursion set theory, the abundance of such objects is determined by the first-crossing distribution of a barrier with height
$\delta_c(z)\approx 1.686$ by Markovian random walks \cite{Shen:2005wd}.
Here, $\delta_c$ is the linear, critical collapse threshold in the spherical collapse approximation.
Consequently, the average, logarithmic number density $\bar n_f(M)$ of filaments of {\it dark matter mass}
$M=f_g^{-1}\mathcal{M}_g\cdot L$, where $L\sim 1\hmpc$ is the proper length of the filament, is given by
\begin{equation}
  \bar n_f(M,z) = \frac{dN_f}{d\ln M} = \frac{\bar\rho_m}{M}\nu f(\nu) \frac{d\ln\sigma}{d\ln M} \;,
\end{equation}
where $\nu(M)\equiv \delta_c/\sigma(M)$ is the peak height, $\sigma(M,z)$ is the root mean square (r.m.s.) variance of the density field linearly
extrapolated to redshift $z$, and the first crossing distribution reads
\begin{equation}
  \nu f(\nu) = \sqrt{\frac{2}{\pi}} \nu\, e^{-\nu^2/2} \;.
\end{equation}
This implies that the fraction of mass in filaments with mass greater than $M$ is
\begin{equation}
  F_f(>M,z) = {\rm Erfc}\!\left(\frac{\nu(M)}{\sqrt{2}}\right) \;.
\end{equation}
Assuming $L=1\hmpc$ as suggested by numerical simulations, $z=3$ filaments with a baryonic mass per unit length $\mathcal{M}_g=10^{10-11}\masslength$
have an abundance in the range $n_f\sim 10^{-3} - 10^{-1}$. Hence, there are ubiquitous in the high-redshift cosmic web. However, their abundance
drops quickly below $10^{-5}$ as soon as $M$ exceeds $10^{13}\hmsun$.

Assuming $A_\text{c}=1$, the critical axion core mass inferred from Eq.~(\ref{eq:core-mass}) should thus be $\mathcal{M}_\text{c,max}\lesssim 0.01 - 0.1$
(in unit of $10^{10}\masslength$), that is, $f_{17}\lesssim 10^{-4}$ for the axion attractive self-interaction to strongly affect low column density
Lyman-$\alpha$ absorbers.

\subsection{Including the gravitational pull from the axion core}
\label{sec:solvingLE}

Eq. (\ref{eq:lanemden}) shows that, when the dark matter source is included, the second derivative $\theta''(\xi)$ becomes more negative near the origin.
In this case, the position $\xi_1$ of the first zero could thus be noticeably smaller than obtained with the homogeneous equation if the dark matter
density is sufficiently large. 
More precisely, we expect that the product $\big\lvert\xi_1\theta'(\xi_1)\big\lvert$ is not much affected since $\theta'(\xi_1)\sim 1/\xi_1$, but the
integral of the {\small HI} profile $\theta^{n\beta}$ (see Eq.~{\ref{eq:NHI}) becomes much smaller.
Therefore, this may eventually yield column densities $N_\text{\tiny HI}$ lower than naively inferred above.

In order to get some quantitative estimate for the impact the axion core on the filamentary gas profile, we ignore the haze of virialized axions and
set $\rho_\text{DM}=\rho_c$, where $\rho_c$ is the density profile of the solitonic solution, Eq.(\ref{rhofilament}).
Here again, we assume a polytropic index $n=1$, for which the homogeneous Lane-Emden equation with initial conditions $\theta(0)=1$ and $\theta'(0)=0$
admits the solution $\theta(\xi)=J_0(\xi)$ \cite{ostriker:1964}.
For $n=1$, the general solution to the inhomogeneous equation Eq.(\ref{eq:lanemden}) is $\theta(\xi)=J_0(\xi)+\theta_p(\xi)$,
where the particular solution $\theta_p(\xi)$ solves the inhomogeneous problem with initial conditions $\theta_p(0)=\theta_p'(0)=0$.
For $0\leq\xi<\xi'$, the Green's function $G(\xi,\xi')$ satisfying $G(0,\xi')=\partial_\xi G(0,\xi')=0$ is trivially $G(\xi,\xi')=0$.
For $\xi'>\xi$, we seek a solution of the form $G(\xi,\xi')=C(\xi')J_0(\xi)+D(\xi') Y_0(\xi)$, where $J_0$ and $Y_0$ are independent solutions to the
homogeneous equation.
Applying the continuity and jump condition at $\xi=\xi'$, the Green's function eventually reads
\begin{equation}
  \label{eq:green}
G(\xi,\xi') = \Theta(\xi-\xi') \frac{J_0(\xi')Y_0(\xi)-J_0(\xi)Y_0(\xi')}{\big(J_1(\xi')Y_0(\xi')-J_0(\xi')Y_1(\xi')\big)} \;.
\end{equation}
The physical interpretation of the Green's function is straightforward: $G(\xi,\xi') = 0$ for $\xi'>\xi$ because Birkhoff's theorem ensure that the gas
profile $\theta(\xi)$ only depends on the dark matter source at $\xi'<\xi$.

The solution to the inhomogeneous Lane-Emden equation thus 
\begin{equation}
  \label{eq:sol}
\theta(\xi) = J_0(\xi) -\frac{1}{\rho_0} \int_0^\infty\!d\xi' \, G(\xi,\xi') \rho_c(\xi') \;,
\end{equation}
and, for our fiducial choice of $m_{22}$ and $f_{17}$, is very sensitive to the normalization $f_\text{c}$ of the axion core - filament mass relation.
In principle, the solution should be refined iteratively from an initial guess because the axion core mass $\mathcal{M}_\text{c}$ depends on the baryonic
mass $\mathcal{M}_g$ of the filament, which is itself a function of $\big\lvert\xi_1\theta'(\xi_1)\big\lvert$.
However, since $\mathcal{M}_\text{c}\propto \mathcal{M}_g^{1/3}$ weakly depends on $\mathcal{M}_g$ and, furthermore,
$\big\lvert\xi_1\theta'(\xi_1)\big\lvert\sim \mathcal{O}(1)$ for the range of parameters considered here, we shall skip this iterative search and simply
set the baryonic mass to $\mathcal{M}_g=3.2\times 10^{10}\Delta \masslength$.
The total dark matter mass can then be inferred upon assuming a gas fraction $f_g$.

Finally, let us emphasize again that, for a decay constant $f_{17}\geq 0.01$ required to match the observed density of dark matter without much fine-tuning
(and a possibly temperature-dependent axion mass \cite{diez/marsh:2017}),
the mass of the solitonic axion core is always orders of magnitude smaller than $M_\text{c,max}$.
Consequently, unless the normalization $A_\text{c}$ is extremely high (i.e. $A_\text{c}\gg 10^3$), which is very unlikely, $R_\text{c,filament}$ is weakly
dependent on $f_{17}$ (see Fig.~\ref{fig:mradius}). Hence, the conclusions drawn here hold regardless of the exact value of the decay constant provided
that $f_{17}\gtrsim 0.01$.

Fig.~\ref{fig:lanemden} shows various solutions spanning the range $1\leq\Delta\leq 10$ (as is appropriate to the mildly nonlinear Lyman-$\alpha$ forest)
for our fiducial axion mass $m_{22}=1$ and decay constant $f_{17}=1$.
The dark matter profile is the Gaussian Eq. (\ref{rhofilament}), with a radius $R_\text{filament}$ given by the stability condition Eq. (\ref{eq:Rfilament}).
The top and bottom panels assume a normalization $A_\text{c}=10$ and $10^3$, respectively.
As a result, the axion solitonic core mass and radius are
\begin{gather}
  \mathcal{M}_\text{c}\sim 10^9\masslength\;, \\
  R_\text{filament}\sim 4\hkpc \nonumber
\end{gather}
for $A_\text{c}=10$, and
\begin{gather}
  \mathcal{M}_\text{c} \sim 2\times 10^{10}\masslength\;, \\
  R_\text{filament} \sim 1\hkpc \nonumber
\end{gather}
for $A_\text{c}=10^3$.
The impact of the axion core is largest at low gas central overdensity $\Delta$ owing to the mass dependence $\mathcal{M}_\text{c}\propto \mathcal{M}_g^{1/3}$
of the axion core - filament mass relation. Whereas, for $A_\text{c}=10$, the axion leaves a small signature on the profile, the latter becomes significantly
more compact for $\Delta\leq 5$ when $A_\text{c}=10^3$.  
In the particular case $\Delta=1$, the {\small HI} column density drops from $N_\text{\tiny HI}\sim 5\times 10^{13}\cmm$ ($A_\text{c}=10$) down to
$\sim 10^{12}\cmm$ ($A_\text{c}=10^3$) while its radius shrinks to $\sim 8\hkpc$.
Since the line column density $f(N_\text{\tiny HI},z)$, Eq.~(\ref{eq:fHIz}), depends on both the filament cross-section and profile, we expect that the
Lyman-$\alpha$ forest should be strongly affected if $A_\text{c}$ is as large as $10^3$.

Overall, this demonstrates that the gravitational pull sourced by the solitonic axion core in the Lane-Emden equation is crucial to our discussion.
Obviously, the question of whether axion self-interactions leave a signature in the Lyman-$\alpha$ forest can only be fully addressed with numerical
simulations, which can also be used to extract the value of $A_\text{c}$.
Nonetheless, we believe our analytic approach provides useful insight into this issue.

\section{Conclusions}
\label{sec:conclusions}
\noindent
In this paper we have taken the first step towards the understanding of the impact of a tiny, but non-vanishing, self-interaction among  ultra-light axions
on the large scale structure of the Universe. 
We have considered axion masses $m=m_{22}\cdot 10^{-22}$ eV and decay constants $f=f_{17}\cdot 10^{17}$ GeV, for which the axions can provide a significant fraction
of the dark matter. Our analytical investigation based on the GP equation and on the breathing mode  has shown that (for $m_{22}=1$) 

\begin{itemize}
\item Spherical axion cores are stable only if their masses are smaller than about $7\times 10^{9} f_{17}\,\hmsun$,
  in agreement with the findings of \cite{Chavanis:2011zi}.
  We extended the analysis of \cite{Chavanis:2011zi} and emphasized that gravity is essential in rendering halos stable below the critical mass.
  With no gravity taken into account, the tiny self-interaction strength would make halos of all masses collapse.
  The potential of  the halo breathing mode (fluctuations of its radius) develops a local minimum which is destroyed for halos above a mass threshold.
  Thus, halos with mass below the cutoff mass are stable against spherical symmetric perturbations in the presence of gravity. 

\item Pancake-like solitonic cores are stable if one restricts oneself to the breathing mode, but they are  unstable against transverse perturbations.
  Our computation emphasizes that, although one-dimensional soliton solutions of the GP
  equation exists, these pancake-like objects do not survive the presence of fluctuations along the transverse directions once they are immersed in a
  higher-dimensional environment.
  Gravity does not seem to alter this conclusion.

\item Axions cores within filaments are stable if their mass per unit length is smaller  than the critical value
\begin{equation}
{\cal M}_\text{max} =  7.0\times 10^{16} f_{17}^2\,\masslength \; .
\end{equation}
Filamentary axion cores with mass ${\cal M}_\text{c}\gtrsim {\cal M}_\text{c,max}$ are unstable and collapse by redshift $z=0$ but, for $f_{17}\gtrsim 0.01$,
they are expected to form inside objects which are so rare that this instability would never be observed.
Therefore, within the approximations made in this paper, the attractive axion self-interaction does not have an impact on the Lyman-$\alpha$ forest unless
$f_{17}\lesssim 10^{-4}$, or the normalization $A_\text{c}$ of the axion core -filament mass relation (relative to its value for spherical halos) is very
high. Notwithstanding, hydrostatic equilibrium considerations suggest that, for $f_{17}\geq 0.01$, axion solitonic cores inside filaments will leave a
detectable impact on the distribution of Lyman-$\alpha$ absorption lines provided that $A_\text{c}\gtrsim 10^2$.
This effect arises from the gravitational pull of the axion core, which affects the gas density profile and estimated {\small HI} column densities of
high-redshift filaments.

\end{itemize}
Notice that at small radii the breathing mode analysis reveals that gravity is always negligible in the corresponding potential.
Furthermore, the existence of critical mass scales for axion halos and filaments should be contrasted to the standard CDM case, for which the global
energetics analysis performed here predicts the existence of solutions at all mass owing to the centrifugal barrier (CDM filaments are indeed observed in
realistic large scale structure CDM simulations, see e.g. \cite{park:1990}).
Therefore, our findings suggests that the cosmic web may look significantly different if the dark matter is a light self-interacting axion rather than a cold,
massive fermion.

Our analysis can be improved in  several ways.
Firstly, we have not discussed transverse perturbations for filaments and halos in the presence of gravity.
It is not unreasonable to expect that they might lead to  instability as well. We leave this analysis for the future.

Secondly, one could ask what is the role of the higher-order terms in the GP equation derived from the axion periodic potential. 
They might indeed provide  a defocusing (repulsion)   needed to support the stability. In fact,  the full axion potential may assist the stability. 
One can for example calculate the internal energy of a filament with the Gaussian  density profile (\ref{rhofilament}).
In this case, we find that the axion self-energy (in the regime where the full axion potential is relevant) is given by 
\begin{align}
  \label{uuu}
 U &=48\zeta_2\pi^2{\cal M}^2  R^2\bigg\{1-\gamma+{\rm Ci}\left(\frac{1}{2\pi R^2}\right) \\
 & \qquad +\log(2\pi R^2)-2\pi R^2 {\rm Si}\left(\frac{1}{2\pi R^2}\right)\bigg\} \nonumber \;,
 \end{align}
where Ci and Si  are the cosine  and sine integral functions and $\gamma$ is the Euler gamma. It is easy to verify that for large $R\gg 1$,
$U\approx \zeta_2{\cal M}^2/R^2$, which is what we found  in \S\ref{sec:filaments}.
Therefore, keeping only the leading $\phi^4$ term from the axion potential in the non-relativistic limit,  we can wrongly conclude that the filaments
are unstable.
However, for $R\ll 1$, the internal energy is given instead in Eq. (\ref{uuu}) and it is straightforward to check that the total energy (the quantum
pressure and the internal energy) now develop a minimum even without the presence of gravity.
Of course, the full potential is relevant only when $\phi\sim f$, that is, during the final stages of the collapse. 
The take home message is that the full axion potential could be relevant for the final fate of halos, pancakes and filaments. 
For instance, it has been recently discovered that even a collapsing condensate can leave behind highly robust soliton configurations in $D=3$ after
collapse \cite{Wieman}.  

Thirdly, although our analytic approach captures the essential features of Lyman-$\alpha$ absorbers in the presence of an axion core, it does not
take into account peculiar velocities, thermal broadening etc. Furthermore, the normalization of the axion core - filament mass is left unconstrained.
It would thus be desirable to investigate these issues further with numerical simulations.
These should also give us insights into the importance of transverse instabilities which we neglected here.

We conclude by mentioning that the non-linear processes induced by the self-interactions might also lead to other unexpected phenomena if the dark matter
is composed by ultra-light axions.
One of them  is the so-called four-wave mixing process observed in experiments \cite{four} where three initial  wave-packets interact non-linearly
to produce a fourth packet.
This phenomenon might be relevant when considering possible regions of over-densities and the relation between the three- and the four-point correlators
of the dark matter.

\acknowledgments
\noindent
We thank  I. Tkachev for discussions and especially D.J.  Frantzeskakis and T.P. Horikis for illuminating interactions on solitons and their stability.
We are also grateful to Mor Rozner for spotting out a numerical error in Eq.(\ref{eq:kI}); to David Marsh for helpful comments on an earlier
  version of this manuscript; and to the anonymous referees for their constructive suggestions.
V.D. acknowledges support by the Israel Science Foundation (grant no. 1395/16). A.K. is partially supported by GGET project 71644/28.4.16.
A.R. is supported by the Swiss National Science Foundation (SNSF), project {\sl Investigating the Nature of Dark Matter}, project number: 200020-159223.

\bibliographystyle{prsty}
\bibliography{references}

\label{lastpage}

\end{document}